\documentclass{aa}
\usepackage{natbib}
\usepackage{graphicx}
\usepackage{txfonts}

\def\nifs{\iso{56}Ni}
\def\cofs{\iso{56}Co}

\def\cm3{cm$^{-3}$}
\def\kms{\mbox{km~s$^{-1}$}}

\def\msun{$M_{\odot}$}

\def\one{\ts {\,\sc i}}
\def\two{\ts {\,\sc ii}}
\def\three{\ts {\,\sc iii}}
\def\four{\ts {\,\sc iv}}

\def\beq{\begin{equation}}
\def\eeq{\end{equation}}

\def\lesssim{\mathrel{\hbox{\rlap{\hbox{\lower4pt\hbox{$\sim$}}}\hbox{$<$}}}}
\def\gtrsim{\mathrel{\hbox{\rlap{\hbox{\lower4pt\hbox{$\sim$}}}\hbox{$>$}}}}

\def\one{{\,\sc i}}
\def\two{{\,\sc ii}}
\def\three{{\,\sc iii}}
\def\four{{\,\sc iv}}

\def\mch{$M_{\rm ch}$}

\def\v1d{{\code{V1D}}}

\def\cmfgen{{\code{CMFGEN}}}

\newcommand{\code}[1]{\texttt{#1}}

\def\ergs{erg\,s$^{-1}$}

\newcommand{\iso}[2]{\ensuremath{^{#1}\rm{#2}}}

\def\aj{AJ}
\def\pasp{PASP}

\def\apj{ApJ}
\def\apjs{ApJS}
\def\apjl{ApJL}
\def\aap{A\&A}

\def\mnras{MNRAS}
\def\nat{Nature}

\begin{document}

 \title{Spectral signatures of H-rich material stripped from a non-degenerate companion by a Type Ia supernova}
 \titlerunning{Stripped material by type Ia SN ejecta}

\author{
 Luc Dessart\inst{\ref{inst1}}
 \and
 Douglas C. Leonard\inst{\ref{inst2}}
 \and
 Jose~L. Prieto\inst{\ref{inst3},\ref{inst4}}
 }

\institute{
Institut d'Astrophysique de Paris, CNRS-Sorbonne Universit\'e, 98 bis boulevard Arago, F-75014 Paris, France.\label{inst1}
\and
 Department of Astronomy, San Diego State University, San Diego, CA 92182-1221, USA.\label{inst2}
\and
N\'ucleo de Astronom\'ia de la Facultad de Ingenier\'ia y Ciencias, Universidad Diego Portales,
Av. Ej\'ercito 441 Santiago, Chile.\label{inst3}
\and
Millennium Institute of Astrophysics, Santiago, Chile.\label{inst4}
  }

   \date{Received; accepted}

  \abstract{
The single-degenerate scenario for Type Ia supernovae should yield metal-rich ejecta that enclose some stripped material from the non-degenerate H-rich companion star.  We present a large grid of non-local thermodynamic equilibrium steady-state radiative transfer calculations for such hybrid ejecta and provide analytical fits for the H$\alpha$ luminosity and equivalent width.  Our set of models covers a range of masses for \nifs\ and the ejecta, for the stripped material ($M_{\rm st}$), and post-explosion epochs from 100 to 300\,d. The brightness contrast between stripped material and metal-rich ejecta challenges the detection of H\one\ and He\one\ lines prior to $\sim$\,100\,d. Intrinsic and extrinsic optical depth effects also influence the radiation emanating from the stripped material. This inner denser region is marginally thick in the continuum and optically thick in all Balmer lines. The overlying metal-rich ejecta blanket the inner regions, completely below about 5000\,\AA, and more sparsely at longer wavelengths. As a consequence, H$\beta$ should not be observed for all values of $M_{\rm st}$ up to at least 300 days, while H$\alpha$ should be observed after $\sim$\,100\,d for all $M_{\rm st} \geq$\,0.01\,\msun. Observational non-detections capable of limiting the H$\alpha$ equivalent width to $<$\,1\,\AA\ set a formal upper limit of $M_{\rm st} < 0.001$\,\msun. This contrasts with  the case of circumstellar-material (CSM) interaction, not subject to external blanketing, which should produce H$\alpha$ and H$\beta$ lines with a strength dependent primarily on CSM density.  We confirm previous analyses that suggest low values of order 0.001\,\msun\ for $M_{\rm st}$ to explain the observations of the two Type Ia supernovae with nebular-phase H$\alpha$ detection, in conflict with the much greater stripped mass predicted by hydrodynamical simulations for the single-degenerate scenario. A more likely solution is the double-degenerate scenario, together with CSM interaction, or enclosed material from a tertiary star in a triple system or from a giant planet.
}

   \maketitle

\section{Introduction}

Unambiguous signatures of the single-degenerate scenario for Type Ia supernovae (SNe Ia) are much desired but hard to secure. One such signature is the identification of material stripped from the companion star by the SN Ia ejecta. Analytic explorations \citep{wheeler_snia_75,chugai_snia_86} and numerical simulations \citep{marietta_snia_hyd_00,pakmor_snia_ms_08,liu_snia_12,pan_snia_12,liu_snia_ha_17} of such a scenario have been performed in recent decades and have yielded similar conclusions. The key signatures are that $0.1-0.5$\,\msun\ of material should systematically be stripped from the non-degenerate companion star, asymmetrically distributed but limited to low velocities below about 1000\,\kms\ in the resulting ejecta. The main radiative signature for this scenario is proposed to be the appearance at late times of a narrow and strong H$\alpha$ line on top of a normal SN Ia spectrum \citep{chugai_snia_86,mattila_01el_ha_05,botyanszki_ia_neb_sd}.  More sophisticated and complete studies of the progenitor evolution do not alter this picture (see, for example, \citealt{liu_snia_ha_17}).

Narrow H$\alpha$ emission has been detected in a number of SNe Ia. In some events (e.g., SNe 2002ic or 2005gj), the detection is early, around bolometric maximum, and compatible with circumstellar-material (CSM) interaction  \citep{hamuy_02ic_03,silverman_ian_13,kotak_02ic_04}. For two noteworthy events, the H$\alpha$ emission has been detected later, at $\sim$\,100\,d past maximum or more (for example, SN\,2018cqj, \citealt{prieto_var_ha_19};  ASASSN-18tb, \citealt{kollmeier_18tb_ha_19,vallely_snian_19}). These events may be explained by either CSM interaction (in particular when the detection occurs earlier, as for ASASSN-18tb as $+39$\,d; \citealt{vallely_snian_19}) or by emission from stripped material from a companion. However, in the latter case, radiative transfer simulations, confirmed by the present work, suggest the stripped material mass $M_{\rm st}$ is on the order of 0.001\,\msun.   This is much less than predicted by all hydrodynamical simulations of this scenario, which suggest that at least 0.1\,\msun\ of material should be stripped (see original work by \citealt{marietta_snia_hyd_00} and the most recent complete study by \citealt{liu_snia_12}).  Another tension for the single-degenerate scenario is that such H$\alpha$ emission is very rarely seen (stringent upper limits on $M_{\rm st}$ are placed instead), even for the most nearby SN Ia events or those with very high quality observations (\citealt{leonard_07}; \citealt{lundqvist_snia_neb_13, lundqvist_snia_neb_15};  \citealt{shappee_snia_ha_13, shappee_snia_ha_18};  \citealt{maguire_snia_ha_16}; \citealt{graham_snia_ha_17}; \citealt{sand_snia_ha_18}; \citealt{holmbo_snia_ha_19}; \citealt{dimitriadis_snia_ha_19}; \citealt{sand_snia_ha_19}; \citealt{tucker_snia_ha_19a, tucker_snia_ha_19}),  while theory predicts that it should instead be systematically detected. In the current framework, this is a major problem for the single-degenerate scenario for SNe Ia. One potential way out of the discrepancy may be to modify the donor star through some mechanism so that the amount of stripped, hydrogen-rich material is vastly reduced.  Such mechanisms could be hydrogen stripping by an optically thick wind so that it has a helium-rich envelope \citep{hachisu_snia_08,pan_snia_10} or a ``spin-up spin-down'' phase that strips the envelope and also leaves a much more compact star \citep{justham_snia_11,hachisu_snia_12}.

In this work, we focus on the emission properties that stripped material will have on the resulting spectrum. So far, only a few studies have been published on the non-local thermodynamic equilibrium (non-LTE) radiative-transfer modeling for SNe Ia with stripped material from a companion star. \citet{mattila_01el_ha_05} and \citet{lundqvist_snia_neb_13,lundqvist_snia_neb_15} use parametrized ejecta structures together with the radiative-transfer code of \citet{kozma_snia_05} to set constraints on the H$\alpha$ emission from the stripped material. Making a number of simplifications for the radiative transfer \citep{botyanszki_ia_neb_17}, and in particular pure optically-thin line emission, \citet{botyanszki_ia_neb_sd} perform similar (i.e., Monte Carlo) simulations but based on physically consistent 3D hydrodynamical simulations of the ejecta--companion interaction. The study of \citet{botyanszki_ia_neb_sd} predicts H$\alpha$ luminosities that appear a hundred to a thousand times greater than those predicted by \citet{mattila_01el_ha_05} for the same stripped material mass and composition. These various studies use the Sobolev approximation, and  treat approximately, or ignore, line overlap and multiple scattering.

In this paper, we build on these previous studies and conduct non-LTE steady-state radiative transfer simulations for SNe Ia with stripped material for a range of post-explosion times, SN Ia ejecta (total mass and \nifs\ mass), as well as a large range of stripped material masses $M_{\rm st}$. The simulations are performed with the standard setup for SN Ia calculations with \cmfgen\ (see, for example, \citealt{HD12,d14_tech,wilk_snia_neb_19}), treat all non-LTE processes, and employ a large model atom. The Sobolev approximation is not used, and line overlap and line blanketing are computed explicitly. This allows for an assessment of the blanketing effects of the metal-rich faster-moving ejecta on the emission from the stripped material, as well as all optical depth effects on continuum and line photons. As in \citet{mattila_01el_ha_05}, we adopt parametrized ejecta structures but also perform tests to evaluate the impact of this shortcoming on our results.

In the next section, we present our numerical approach. Section~\ref{sect_ddc15} studies in detail the results for a reference case, chosen to be a standard Chandrasekhar-mass SN Ia model (i.e., with 0.51\,\msun\ of \nifs) with 0.18\,\msun\ of stripped material. Section~\ref{sect_ha_appearance} elaborates on the processes that control the appearance of H$\alpha$. Section~\ref{sect_gun} presents one clue  for the identification of stripped material. Section~\ref{sect_offset} evaluates the limitations of our approach for the adopted structure of the stripped material. We also discuss the predicted signatures from He\one\ lines in our simulations in Sect.~\ref{sect_he1}. Section~\ref{sect_grid} presents results for the whole grid of simulations performed in this study, encompassing five epochs from 100 to 300\,d, Chandrasekhar as well as sub-Chandrasekhar mass ejecta, \nifs\ masses from about 0.1 to 0.9\,\msun, and stripped material masses from 10$^{-5}$ up to 0.5\,\msun. Section~\ref{sect_ew} emphasizes the distinction between line luminosity and line equivalent width and the relevance for the detection of H$\alpha$ from stripped material in SN Ia spectra. Section~\ref{sect_conc} presents our conclusions.

\section{Numerical approach}

  We investigated  the radiative signatures of stripped material in the innermost layers of SN Ia ejecta using a parametrized approach. For the SN Ia ejecta, rich in iron-group elements and H deficient, we adopted the 1D ejecta structure and composition of the delayed-detonation models DDC0, DDC15, and DDC25 from \citet{b13}. To test the influence of the ejecta mass, we also included the sub-Chandrasekhar mass model SCH3p5 from \citet{blondin_wlr_17}. In the same model order, the \nifs\ mass prior to any decay is 0.86, 0.51, 0.12, and 0.30\,\msun, and the ejecta mass is 1.41, 1.41, 1.41, and 0.98\,\msun\ (Table~\ref{tab_prop}).  The simulations of \citet{mattila_01el_ha_05} and \citet{botyanszki_ia_neb_sd}, to which we will compare our results, are based on the pure-deflagration W7 model of \citet{nomoto_w7_84}. This model is very similar to our model DDC15 below an ejecta velocity of about 10000\,\kms\ (see discussion in \citealt{blondin_15_02bo}). Since this is the region that dominates the emission at nebular times, the relevant metal-rich part of our DDC15 model is analogous to that used in those two studies.

\begin{table}[htbp]
   \centering
\caption{Properties for the SN Ia ejecta used in this study.}
   \begin{tabular}{l|cc}
      \hline
     Model    & $M_{\rm ej}$&    M(\nifs) \\
                      &  [\msun] &  [\msun] \\
\hline
DDC0 &   1.41 &   0.86 \\
DDC15 & 1.41  &  0.51  \\
DDC25 &  1.41 &  0.12  \\
SCH3p5  & 0.98  &  0.30  \\
\hline
   \end{tabular}
   \label{tab_prop}
\end{table}

  For the stripped material, we used a similar approach to \citet{mattila_01el_ha_05} and prescribed a mass (in the range from $\sim$\,10$^{-5}$ up to $\sim$\,0.5\,\msun) and defined the maximum velocity that limits its extension in space (typically 1000\,\kms). We assumed homologous expansion for both regions so there is a direct correspondence between velocity $V$ and radius $R$ for a given SN age $t$ through $R=Vt$. The case of a very low mass of stripped material (i.e., 10$^{-5}$\,\msun) was used to gauge the impact of the stripped material in other models. Because these models have the same setup (grid, model atoms etc), flux subtraction can be used to reveal subtle offsets associated with the emission from the stripped material. This helps, for example, to reveal a global but weak flux offset, or the putative presence of a weak H$\alpha$ line, in particular when it first appears.

\begin{figure}
\includegraphics[width=\hsize]{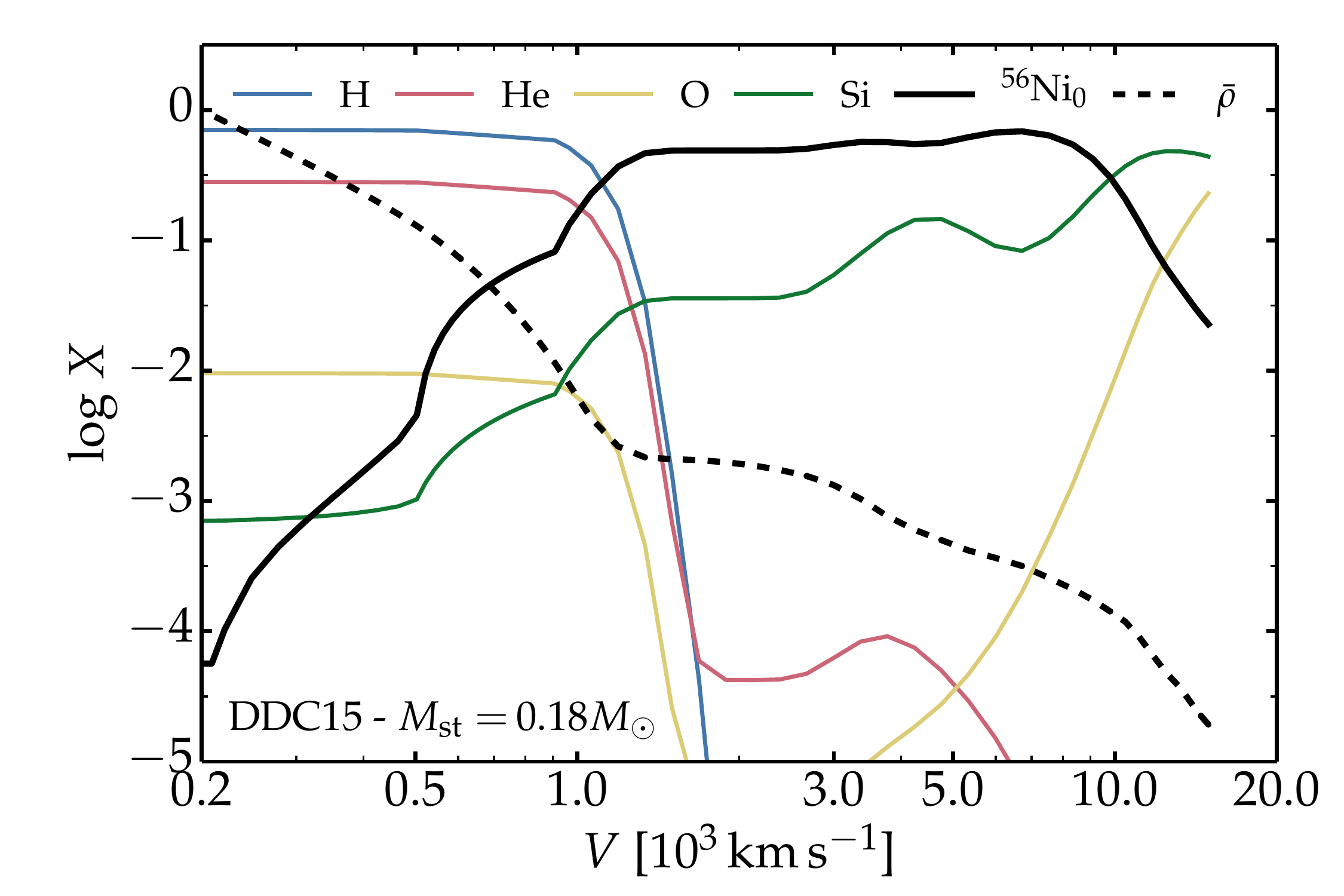}
\vspace{-0.5cm}
\caption{Composition for a selection of elements and normalized density profile (dashed line) in the DDC15 model in which 0.18\,\msun\ of H-rich stripped material has been introduced below 1000\,\kms. The x-axis is shown in logarithmic space to better reveal the innermost regions and the junction between the stripped material (i.e., rich in H and He) and the SN Ia ejecta (i.e., rich in Si and \nifs).
\label{fig_DDC15_ejecta_init}
}
\end{figure}

To setup a model with stripped material, we took the SN Ia ejecta, cut out the inner regions below a chosen velocity limit (typically 1000\,\kms) and replaced what used to be metal-rich material with some H-rich material with a metal composition at the solar value. We also reset the density in this inner region so that it varies as $1/V^2$ and yields the desired mass for the stripped material. The ejecta structure and composition of the SN Ia ejecta are left untouched above that velocity limit. To avoid sharp variations at this new interface, we applied a gaussian smoothing to the density and composition. A representative ejecta structure resulting from this procedure is shown in Fig.~\ref{fig_DDC15_ejecta_init} for the SN Ia model DDC15 with 0.18\,\msun\ of stripped material. Other models varying in SN Ia ejecta composition (DDC0, DDC25), mass (SCH3p5), or in the mass of stripped material, have a similar structure but shifted up and down from the profiles shown in Fig.~\ref{fig_DDC15_ejecta_init}.

This setup is technically unphysical but operationally viable (see discussion in Sect.~\ref{sect_offset} where a comparison for two different stripped-material configurations is presented). The unphysical aspect arises since a real mass-stripping will yield material offset from the center of the explosion, asymmetrically distributed, and yielding line profiles whose width and centroid will be directionally dependent.  This study remains worthwhile since the detectability at late times will not be affected by such considerations.

For the radiative transfer, we used the same technique as described in our SN Ia calculations (see for example \citealt{d14_tech}). However, the model atom was extended to include H and He. We included the following metal species:   C, O, Ne, Na, Mg, Al, Si, S, Ar, Ca, Sc, Ti, V, Cl, K, Cr, Mn, Fe, Co, and Ni.  We included the following ions: H\one, He\one\,--\two, C\one\,--\two, O\one\,--\two, Ne\one\,--\two, Na\one, Mg\two\,--\three, Al\two\,--\three, Si\two\,--\three, S\two\,--\three, Ar\one\,--\three, Ca\two\,--\three, Sc\two\,--\three, Ti\two\,--\three, V\one, Cl\four, K\three, Cr\two\,--\four, Mn\two\,--\three, Fe\one\,--\four, Co\two\,--\four, and Ni\two\,--\four. The model atom used for all these atoms and ions is as described by \citet{DH11_2p} and in the appendix A of \citet{d14_tech}. The large model atom was used for Co\two\ and Co\three. This implies that our calculations treat 1.66 million bound-bound transitions. We also included the same set of decay routes as in \citet{d14_tech} and described in Appendix B. Thus, we included the two-step decay chain of \nifs, as well as other two-step and one-step decay chains (see \citealt{d14_tech} for details and decay constants used). Unlike in \citet{d14_tech}, wherein a Monte Carlo transport solver was used for the $\gamma$ rays, we here employ a simple pure-absorption model (with $\kappa_\gamma=0.06\,Y_{\rm e}$\,cm$^{2}$\,g$^{-1}$, where $Y_{\rm e}$ is the electron fraction) for the computation of the $\gamma$-ray energy deposition \citep{swartz_gray_95,wilk_gammaray_19}.

Since we focus on late times, the very high velocity ejecta are optically thin. We thus reduced the maximum velocity of our grid to 15000\,\kms. In contrast, the stripped material at low velocity is still very dense at $100-300$\,d for the larger values of $M_{\rm st}$. This material has a continuum optical depth near unity, while Balmer lines remain optically thick, even at 300\,d. The configurations explored here are thus distinct, from the point of view of the radiative transfer, from those of standard SNe Ia at a few hundred days. We will discuss these aspects in detail in the Results section below.

Unlike all previous simulations of SNe Ia with stripped material, which employ a Monte Carlo technique and use the Sobolev approximation, we treated the finite intrinsic width of lines and allowed for line overlap \citep{hm98}. We also introduced a turbulent velocity of 10\,\kms. This value corresponds approximately to the thermal velocity of H atoms at 6000\,K, but overestimates that of metals by a factor of a few. However, this is probably more physical for the radiative transfer (especially for H) than the zero intrinsic line width assumed in the Sobolev approximation.

\citet{botyanszki_ia_neb_sd} argued that some of the differences between their results and those of \citet{mattila_01el_ha_05} stemmed from the higher physical consistency of their approach. There is more than a decade between these two studies, which differ in many ways other than the adopted hydrodynamical structure (one problem is that there is little detail on the radiative transfer and its limitations in either of these studies). \citet{botyanszki_ia_neb_sd} do not compare their results to an equivalent 1D setup. The asymmetric distribution of the stripped material causes a velocity shift of the H$\alpha$ line but it is not clear that it affects the total line flux, especially since they assume optically thin line formation. Furthermore, the hydrodynamical simulations of \citet{botyanszki_ia_neb_sd} predict 0.1 to 0.5\,\msun\ of material stripped from the companion, as in previous studies. For lower masses of stripped material, their simulations are scaled-down versions and thus no longer physically consistent. It is not clear in this context that such adopted configurations are superior to the adopted ejecta of \citet{mattila_01el_ha_05}.

Although our work is based on parametrized 1D ejecta, we investigated the effect of a different distribution  for the stripped material, by shifting it to larger velocities (outer bound at 2000 rather than 1000\,\kms; see Sect.~\ref{sect_offset}). A relevant point to consider is what level of accuracy is needed to contribute to the current debate. Inferred masses of stripped material for ASASSN-18tb and SN2018\,cqj suggest very low values of about  0.001\,\msun\ \citep{kollmeier_18tb_ha_19,prieto_var_ha_19}. Such low values are incompatible with the $0.1-0.5$\,\msun\ stripped material masses expected from the standard single-degenerate scenario for SNe Ia, and this holds whether one uses the model results of \citet{mattila_01el_ha_05} or those of \citet{botyanszki_ia_neb_sd}.

\section{Study of the ``standard'' \mch\ SN Ia model with stripped material}
\label{sect_ddc15}

\begin{figure*}
\includegraphics[width=0.5\hsize]{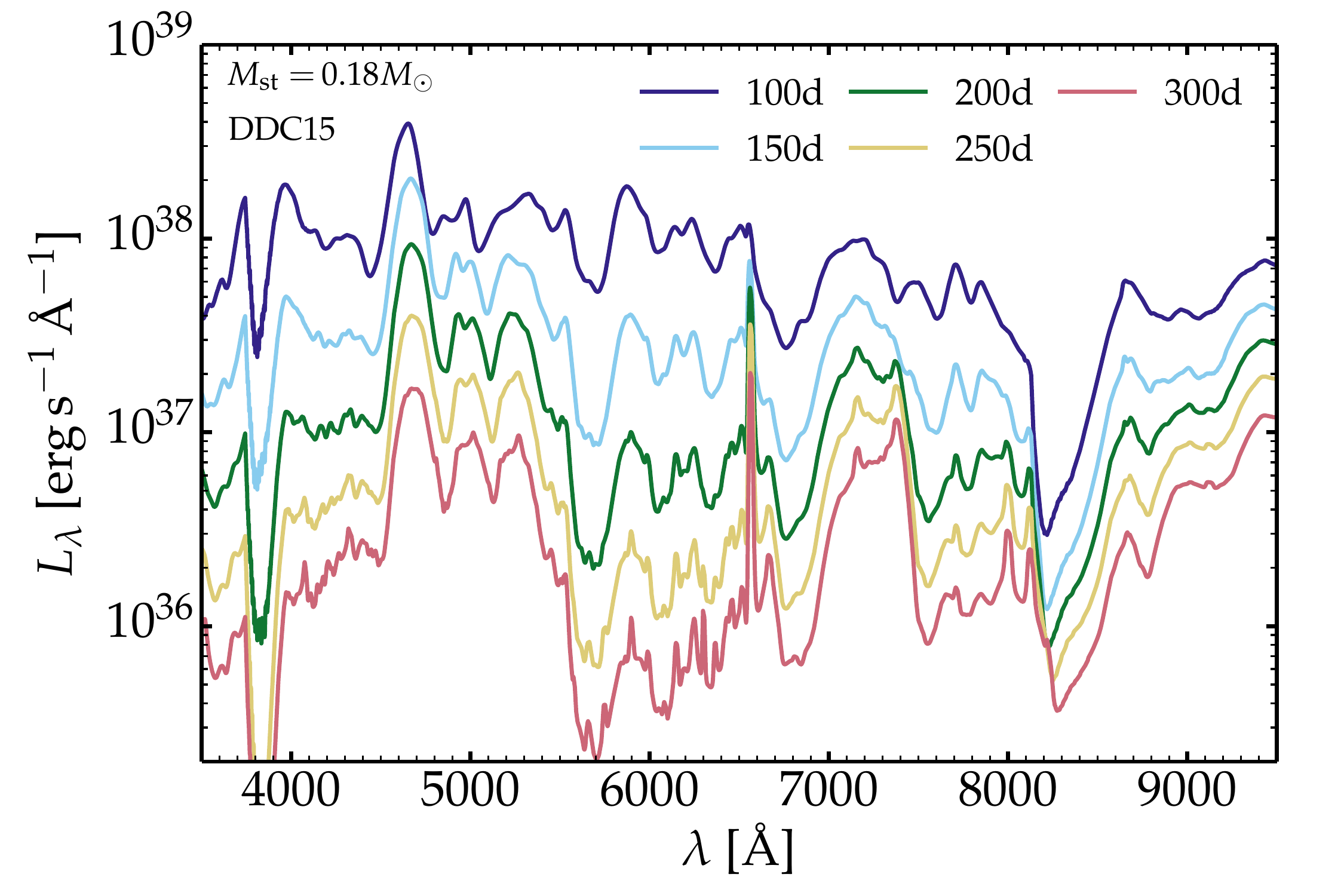}
\includegraphics[width=0.5\hsize]{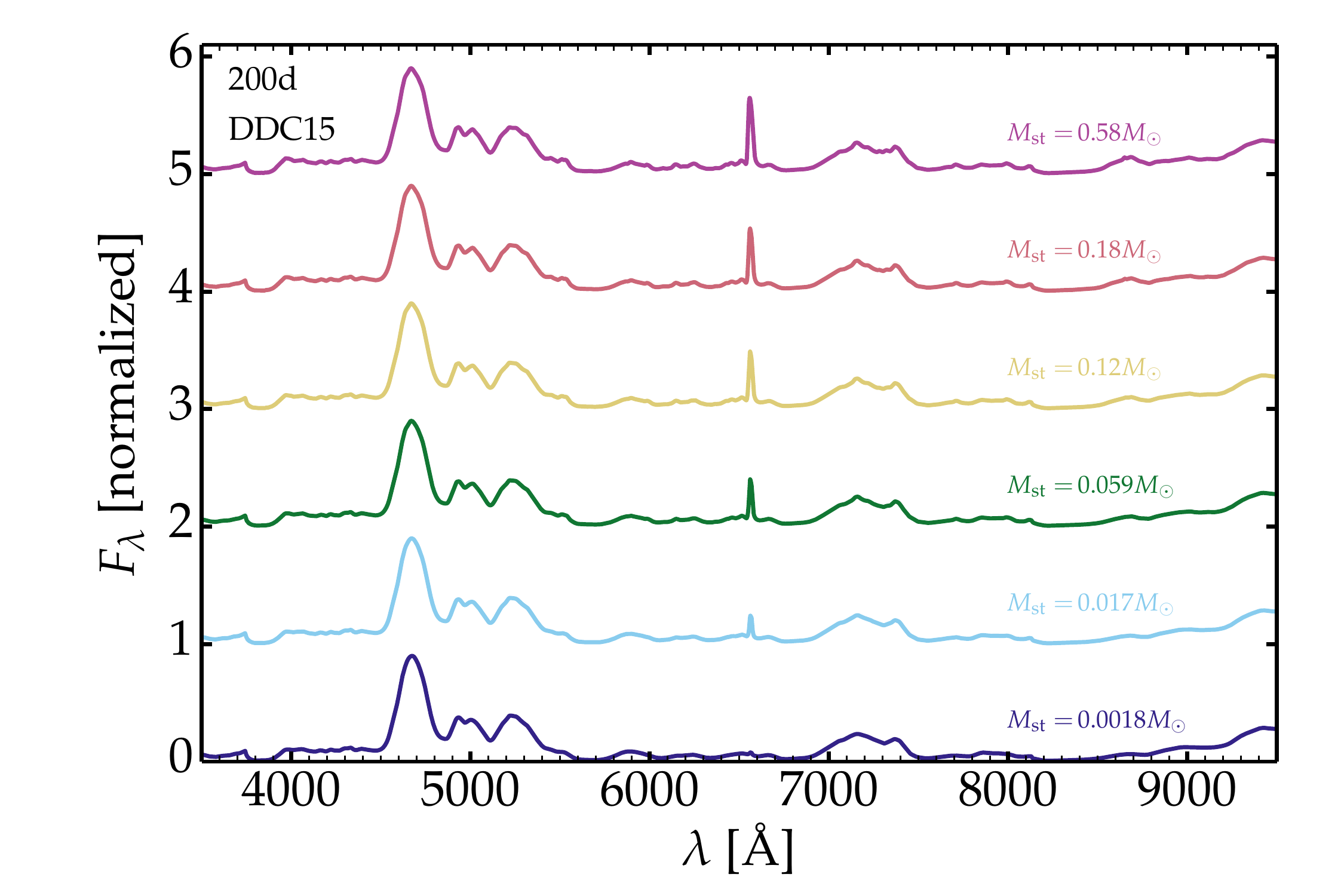}
\caption{Left: Multiepoch spectra (at 100, 150, 200, 250, and 300\,d after explosion) for the SN Ia model DDC15 with a stripped-material mass $M_{\rm st}$ of 0.18\,\msun.  Right: Same as left, but for the model DDC15 at 200\,d and for a range of stripped material mass of 0.0018, 0.017, 0.059, 0.12, 0.18, and 0.58\,\msun. In a qualitative sense, time and $M_{\rm st}$ have a similar impact on the optical spectrum, with a strengthening of the H$\alpha$ line relative to other spectral features with either time or $M_{\rm st}$. Quantitatively, the correspondence does not hold since the bolometric luminosity changes significantly with time but only weakly with $M_{\rm st}$ at a given time.
\label{fig_spec_evol_DDC15}
}
\end{figure*}

The left panel of Fig.~\ref{fig_spec_evol_DDC15} shows the results from our simulations for the DDC15 model, which is representative of a standard SN Ia with its \nifs\ mass of 0.51\,\msun\ (other models are discussed in subsequent sections) with $M_{\rm st}=$\,0.18\,\msun. The main signature associated with the stripped material is the narrow H$\alpha$ line. This is seen as a very weak feature at 100\,d that progressively becomes the strongest line in the optical at 300\,d, exceeding the strength of Fe\three\,4658\,\AA, which is produced by the outer, faster moving metal-rich ejecta. We will not make a comparison to this Fe\three\ line  in the rest of the paper because its strength varies with \nifs\ mass, the ionization level, clumping, and other factors, and it is thus not a robust physical probe. The H$\alpha$ line width never changes because the H-rich material is by design confined to the innermost regions, below about 1000\,\kms. This is done to reflect the results from multi-D hydrodynamical simulations (\citealt{marietta_snia_hyd_00}). The luminosity decrease occurs at all wavelengths but more slowly for the H$\alpha$ line, which eventually becomes the strongest optical line.

The right panel of Fig.~\ref{fig_spec_evol_DDC15} shows the normalized flux for model DDC15 at 200\,d but now for a range of stripped material masses covering from 0.0018 up to 0.58\,\msun. Qualitatively, the evolution with increasing $M_{\rm st}$ is similar to that in time for a fixed value of  $M_{\rm st}$: the H$\alpha$ line strengthens relative to the rest of the spectrum. However, the changes are now roughly at constant luminosity so only the H$\alpha$ line changes with varying $M_{\rm st}$ (this is the reason for plotting the normalized flux). The luminosity is only roughly constant because at a given time a greater decay power is absorbed for increasing stripped material mass (see Sect.~\ref{sect_eabs_ha}).

\begin{figure}
\includegraphics[width=\hsize]{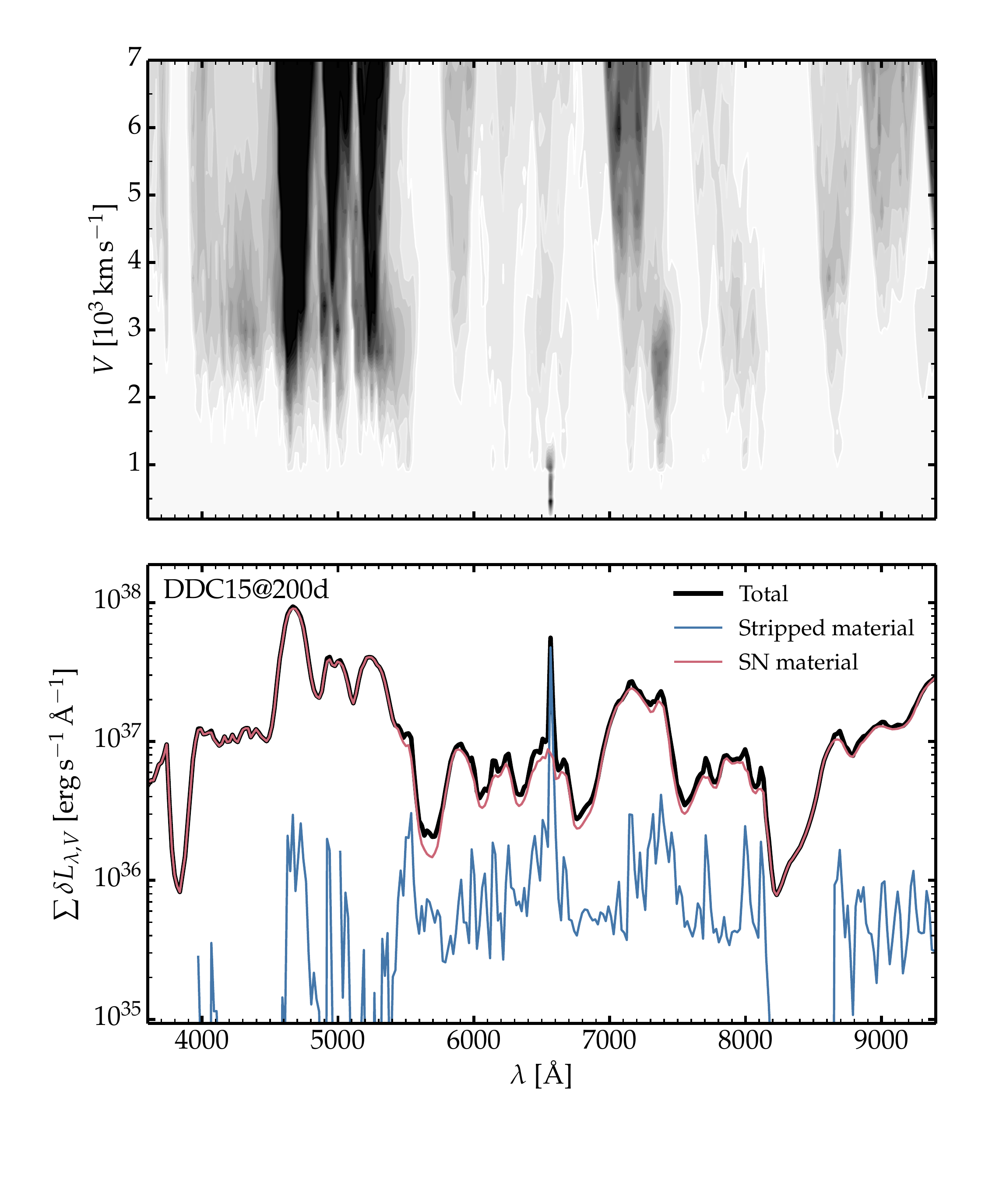}
\vspace{-1.5cm}
\caption{Illustration of the ejecta regions (in velocity space) at the origin of the optical flux for model DDC15 with $M_{\rm st}=$\,0.18\,\msun\ at 200\,d. The top panel shows a gray linear scale of the (emergent) luminosity $\delta L_{\lambda,V}$ while the bottom panel shows the contributions (ordinate in logarithmic space) for the full ejecta and those from the stripped material (blue) or the metal-rich ejecta (red). As is evident from these plots, the predominant emerging feature due to stripped material of the companion star at optical wavelengths is H$\alpha$.
\label{fig_dfr}
}
\end{figure}
\begin{figure}
\includegraphics[width=\hsize]{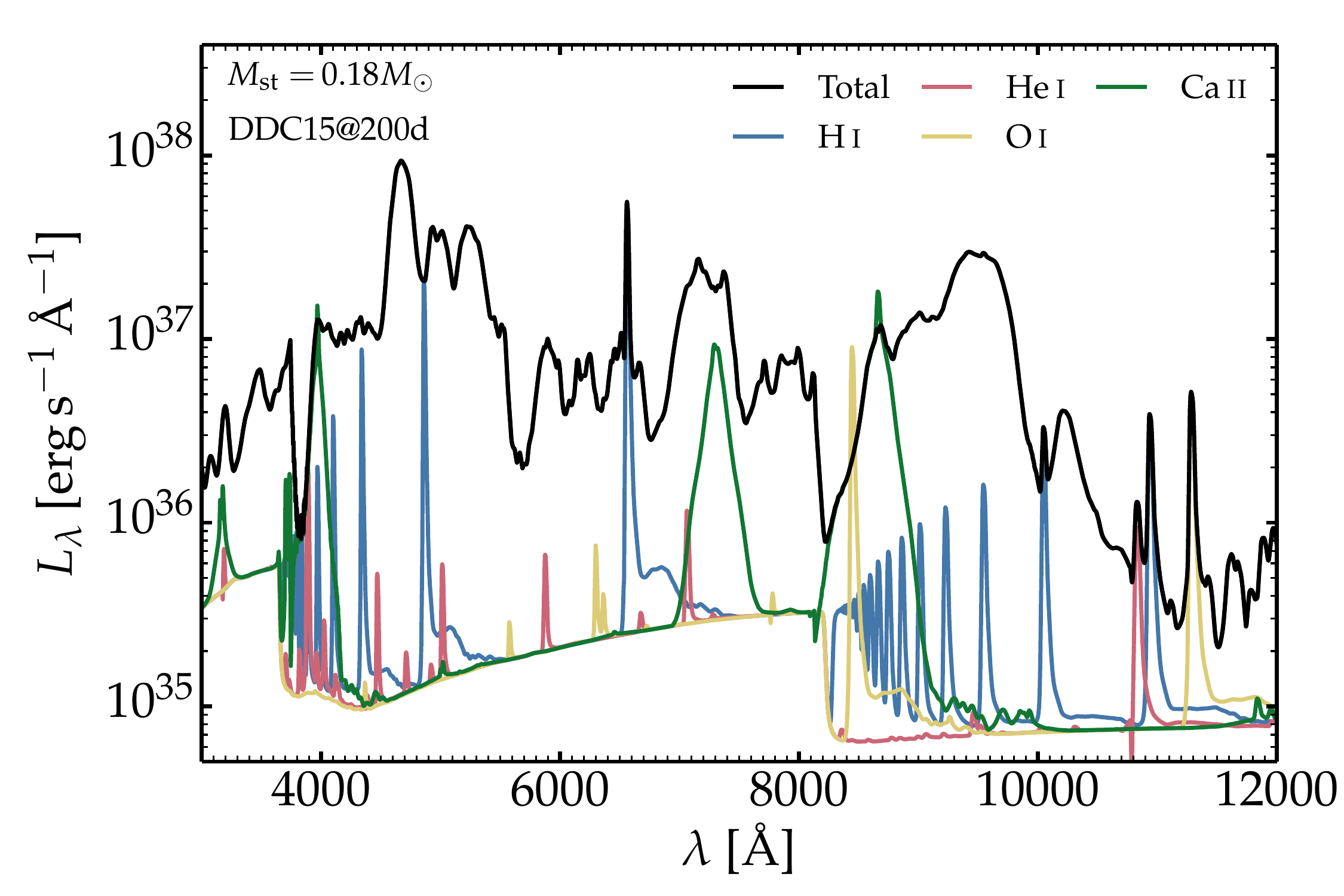}
\vspace{-0.5cm}
\caption{Synthetic spectra for the SN Ia model DDC15 with 0.18\,\msun\ of stripped material. We show the total emergent luminosity $L_\lambda$ as well as that associated with the continuum and bound-bound transition of selected ions (H\one, He\one, O\one, and Ca\two). The calculations for the latter are done by assuming that only the selected ion contributes (together with all continuum processes) and thus ignore the impact of other ions on that selected ion (such as metal-line blanketing on the H\one\ lines arising from the innermost regions associated with the stripped material). The other three models show a similar behavior to that shown here for model DDC15. Hence, while the stripped, low-velocity material is expected to radiate quite strongly at many wavelengths, the primary observable feature at optical wavelengths is due to H$\alpha$.
\label{fig_indiv_spec}
}
\end{figure}

The physical interpretation for these results has been discussed in the past \citep{mattila_01el_ha_05,botyanszki_ia_neb_sd}, although results for multiple epochs have not been shown before. The power source is \cofs\ decay. Of the total decay power absorbed, positrons (whose energy is assumed to be deposited locally) contribute 15\% at 100\,d and this contribution grows roughly linearly in time to 53\% at 300\,d. However, this positron contribution exclusively benefits the metal-rich ejecta, from which the $\gamma$-rays escape increasingly with time. For the stripped material, the situation is very different. Being free of \nifs\ (and \cofs), it can only be powered by the non-local deposition of $\gamma$-ray energy, which is not negligible because of the high density of this H-rich material (this efficiency is also boosted because the opacity to $\gamma$-rays is greater for H-rich material owing to its larger electron fraction, which is about 0.8 rather than 0.5 for symmetric nuclear matter). So, as time passes from 100 to 300\,d, the metal-rich ejecta receive a decreasing fraction of the total decay power absorbed, leading to a relative strengthening of the emission from the stripped material and in particular H$\alpha$ (we find that this line radiates typically 10\% of the decay power absorbed by the stripped material; see Sect.~\ref{sect_eabs_ha}). For an increased stripped material mass (see example in the right panel of Fig.~\ref{fig_spec_evol_DDC15}), more decay power is absorbed and the H$\alpha$ line flux increases relative to the emission from the metal-rich ejecta.

At 200\,d for model DDC15, H$\alpha$ is seen unambiguously for $M_{\rm st}$ greater than about 0.01\,\msun, and is essentially impossible to detect for $M_{\rm st}$ below 0.001\,\msun. An important property of this stripped material is that if H$\alpha$ can be detected around 100\,d, it should be detected for months thereafter since its relative strength to metal lines like Fe\three\,4658\,\AA\ grows in time. A disappearance at late times is not expected in the context of stripped material from a non-degenerate companion. It could be caused by an insufficient decay power absorbed by the stripped material, as may occur if $M_{\rm st}$ were low or very low, but this should compromise the H$\alpha$ detectability at early times as well.

Figure~\ref{fig_dfr} shows how the different ejecta regions (indicated here in velocity space) contribute to the emergent luminosity (or flux) at a given wavelength. This illustration is exact in the sense that the sum of the flux contributions at a given wavelength give the total emergent flux at that wavelength. The stripped material contributes through the strong H$\alpha$ line, as well as some weak and narrow metal lines. The continuum flux is so weak that even weak lines supersede it, with the exception of the spectral region around $6700-7000$\,\AA\ (see also \citealt{chugai_snia_86}). The metal-line emission from the ejecta appears as broad and strong lines (mostly from Fe\two, Fe\three, and Ca\two). For Ca\two\,7300\,\AA\ and some Fe\two\ lines, there is both a broad component from the metal-rich ejecta and a narrow component from the H-rich stripped material (in which all metal mass fractions are at the solar value).

Figure~\ref{fig_indiv_spec} gives a complementary perspective to that  of Fig.~\ref{fig_dfr}. It shows the flux that a given ion would produce in the absence of other ions. This is more artificial than the information presented in Fig.~\ref{fig_dfr} because it ignores any cross-talk between ions, line overlap, and related optical depth effects. By identifying the differences between the full spectrum and the intrinsic contribution of each ion, however, one can assess the importance of optical depth effects (suggested early on by \citealt{chugai_snia_86} and later by \citealt{leonard_07}). These optical depth effects are unambiguously present since the individual ion spectra are rich in many narrow lines, yet none but H$\alpha$ is obviously seen in the total optical flux. The continuum level is 10 to 100 times weaker than the total flux in the emergent spectrum, with an offset that decreases towards longer optical wavelengths.

Some lines suffer from overlap with strong line emission from the metal-rich fast moving ejecta. This strong brightness contrast compromises the identification of the contribution from the stripped material. Other lines suffer from strong attenuation by the metal rich regions, like H$\beta$. In that case, the metal-rich ejecta act as a ``curtain'' that blocks the incoming radiation from the inner regions, as suggested by \citet{leonard_07}. In terms of detectability in the emergent spectrum, the lines of interest sit in regions of lower opacity, and include H$\alpha$, He\one\,1.083\,$\mu$m, H\one\,1.093\,$\mu$m, and O\one\,1.129\,$\mu$m and thus appear as strong narrow features. The resulting emission in the total emergent spectrum is roughly the sum of the individual contributions at the corresponding wavelength. The appearance of narrow lines is in part fortuitous because it arises from the non-uniform distribution of line opacity with wavelength. Overall, the region below 5000\,\AA\ remains thick until late times because of metal-line blanketing, even though the continuum optical depth of these regions is negligible. Consequently, lines like H$\beta$ are attenuated persistently.

Such optical depth effects -- and the associated absence of line features in the resulting spectrum -- have not been treated or discussed by prior studies since strong, narrow features are seen in the predicted synthetic spectrum (see, for example, \citealt{botyanszki_ia_neb_sd}).  This may lead to artificially restrictive limits on the stripped material based on the non-existence of these lines in the observed spectrum (e.g., \citealt{tucker_snia_ha_19a}).

 We note that there are weak bumps on the red side of the strongest lines in the individual ion spectrum calculations (see, for example, the blue curve for H\one\ lines in Fig.~\ref{fig_indiv_spec}). The electron scattering optical depth is below, but close to, unity for the metal-rich ejecta, and this is enough to cause one scattering with a free electron at a large velocity of many thousands of kilometers per second. The associated redshift yields an excess flux between about 10000 and 20000\,\kms\ on the red side of the lines. This is compatible with the ejecta kinematics. The effect cannot be caused by other lines since the H$\beta$ and H$\alpha$ profiles can be overlaid nearly exactly (after a flux scaling). These bumps weaken and move to lower velocities (toward line center) with increasing time, being nearly absent at 300\,d. All these properties are compatible with an electron-scattering origin.

\begin{figure*}
\includegraphics[width=\hsize]{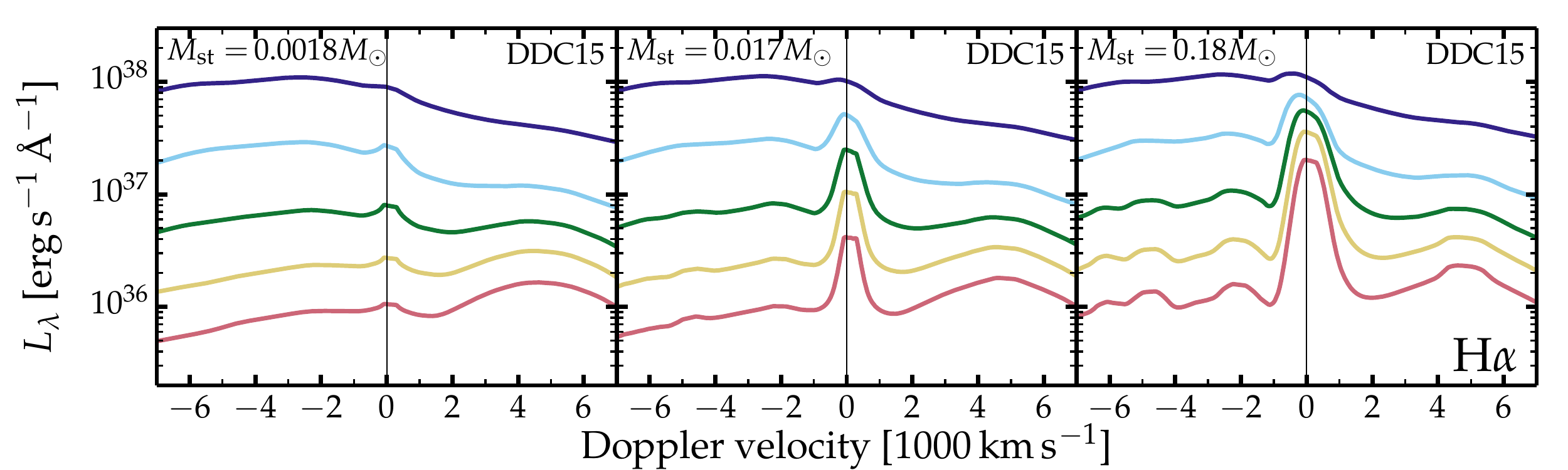}
\includegraphics[width=\hsize]{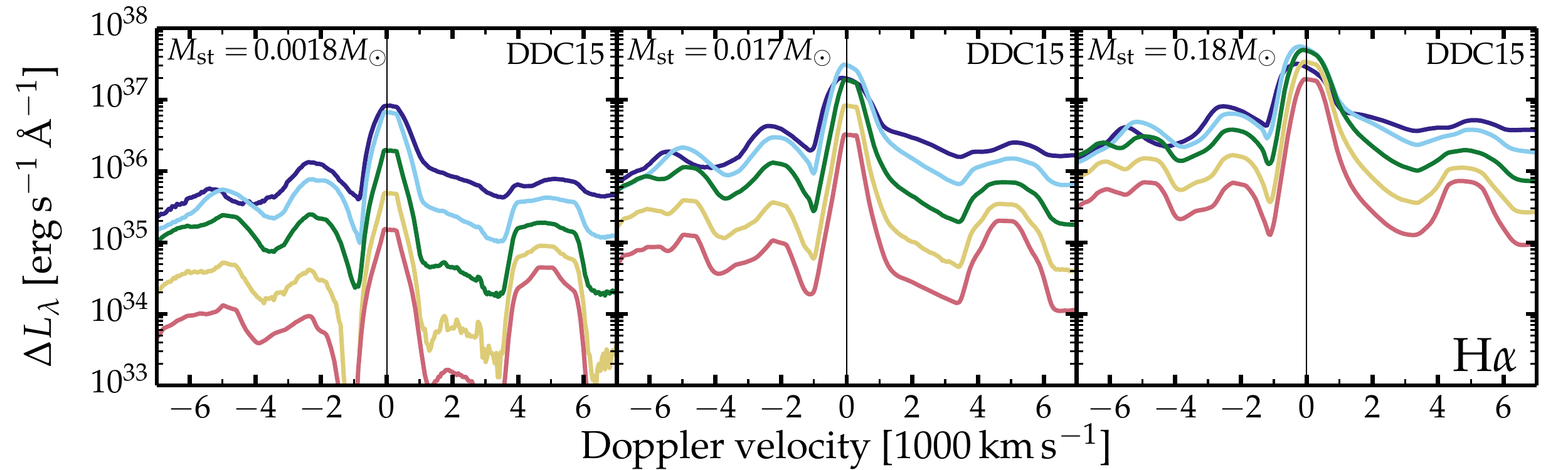}
\caption{Top: Emergent luminosity in the H$\alpha$ region and multiple epochs for model DDC15 with $M_{\rm st}$ of 0.0018, 0.017, and 0.18\,\msun\ from left to right. The color coding indicates epochs 100, 150, 200, 250, and 300\,d after explosion (time increases from the upper curve to the lower curve; see also left panel of Fig.~\ref{fig_spec_evol_DDC15}). Bottom: Same as top, but we now subtract the luminosity arising from the model with $M_{\rm st}=10^{-5}$\,\msun, which has the effect of erasing the luminosity contribution from the metal-rich, SN Ia ejecta. The stripped material contributes not just through H$\alpha$ emission, but also weak emission from other lines (e.g., Fe\two) and some weak background continuum.
\label{fig_muti_panel_Ha_DDC15}
}
\end{figure*}

\begin{figure}
\includegraphics[width=\hsize]{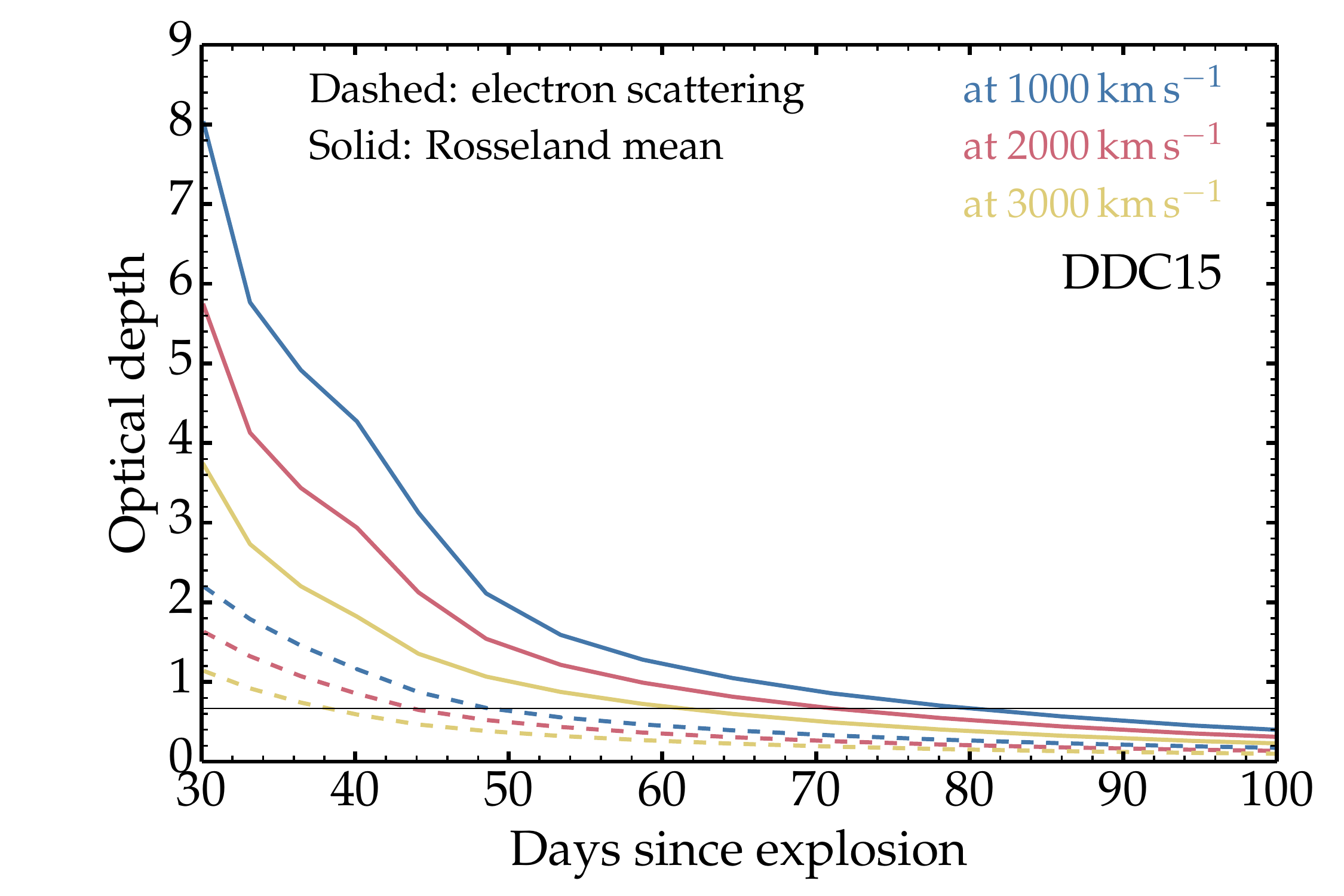}
\vspace{-0.5cm}
\caption{Evolution of the Rosseland-mean (solid) and electron scattering (dashed) optical depth down to the ejecta regions at 1000, 2000, and 3000\,\kms\ for model DDC15 and shown between 30 and 100\,d after explosion. The optical depth of $2/3$ representing the ``photosphere" is shown as a thin black line.
\label{fig_ddc15_tau_at_vel}
}
\end{figure}

\begin{figure}
\includegraphics[width=\hsize]{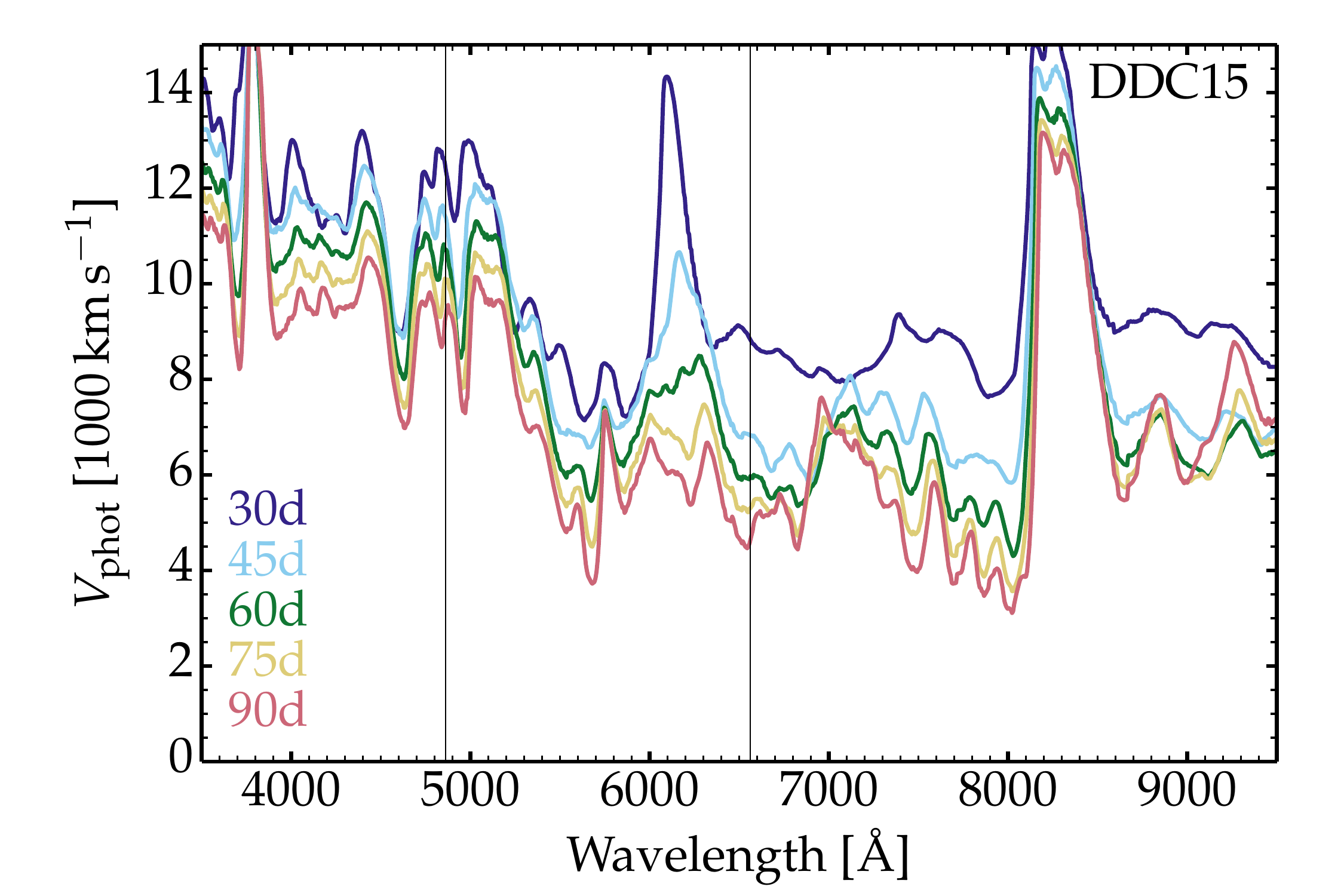}
\vspace{-0.5cm}
\caption{Evolution of the photospheric velocity from 30 to 90\,d after explosion in model DDC15. The two vertical lines give the location of H$\alpha$ and H$\beta$. This computation includes the effects of all opacity sources and is thus more relevant than the Rosseland-mean (because not very physical at low optical depth) or the electron-scattering (because it strongly underestimates the true opacity) optical depth shown in Fig.~\ref{fig_ddc15_tau_at_vel}. This figure shows that H$\beta$ sits in a region of strong blanketing (since photons essentially decouple at $9000-12000$\,\kms\ at the times depicted) while H$\alpha$ sits in one of the most transparent regions of the optical range.
\label{fig_vphot}
}
\end{figure}

\section{Appearance of H$\alpha$ and optical depth effects}
\label{sect_ha_appearance}

   In the simulation DDC15 with $M_{\rm st}=$\,0.18\,\msun\ presented above, the H$\alpha$ line grows from being very weak relative to the rest of the spectrum at 100\,d to becoming the strongest line (in terms of peak flux rather than line flux or equivalent width; see Sect.~\ref{sect_ew})  in the optical at 300\,d. This is shown more clearly in Fig.~\ref{fig_muti_panel_Ha_DDC15} where we isolate the H$\alpha$ region. When inspecting the luminosity (top row), H$\alpha$ is a small bump at 100\,d. Its luminosity is very large but so is that of the metal-rich ejecta so the contrast between H$\alpha$ and the overlapping Fe\two\ lines is small. To better reveal the presence of H$\alpha$, we plot in the bottom row the difference in luminosity between a given model and its counterpart in which $M_{\rm st}=10^{-5}$\,\msun. The H$\alpha$ line is then unambiguously seen, even for $M_{\rm st}=0.0018$\,\msun.

  At early times, the small luminosity contrast between metal-rich regions and the stripped material is one limitation to identifying H$\alpha$. At such times, the $\gamma$-ray escape from the metal-rich regions is moderate so these regions outshine the inner H-rich ejecta. The contribution is more extended so both the regions rich in iron-group and intermediate-mass elements contribute, which leads to a more widespread presence of lines in the optical. With time, the metal-rich region traps less decay power, and the emission is biased inward in favor of the iron-rich regions, so holes appear in the continuum-free spectrum.

  The other limitation is optical depth. There are contributions, of different magnitudes, from electron scattering and lines, and from the metal-rich SN Ia ejecta as well as from the H-rich stripped material. Figure~\ref{fig_ddc15_tau_at_vel} shows the evolution of the electron-scattering and Rosseland mean optical depth at 1000, 2000, and 3000\,\kms\ in the SN Ia ejecta of model DDC15 from 30 to 100\,d after explosion.\footnote{We use here the time sequence for model DDC15 \citep{d14_pddel}, which ignores any stripped material from a non-degenerate companion. Since we focus on the optical depth of the SN Ia ejecta overlying the putative stripped material located below 1000\,\kms, these simulations are suitable for estimating the optical depth of the metal-rich ejecta in the time range 30 to 100\,d.} Because of the significant impact of lines, the latter is well above the former, although one needs to bear in mind that the Rosseland mean is not an ideal representation of opacity in optically thin regions. Nonetheless, one sees that the conditions are only moderately optically thin even at 100\,d and that they are optically thick at 50\,d. Because of expansion, electron scattering redistributes the line flux but does not destroy photons, so this may not quench line emission from the inner stripped material. Metal-line blanketing is in part absorptive but the non-uniform distribution of lines implies that even for optically thick conditions according to the Rosseland mean, some spectral regions may remain opacity holes. This is reflected by the strong variation of the photospheric velocity with wavelength and time (Fig.~\ref{fig_vphot}). One should bear in mind that the photosphere is not an opaque wall below which no photons escape. Instead, it represents the layer of ``median'' emission, in other words about 50\% of the emerging photons come from below that layer and the rest from above. So, ``opaque" regions (i.e., those located at an optical depth greater than unity) do contribute to the emergent radiation. Hence, Fig.~\ref{fig_vphot} is merely illustrative of the non-uniform trapping efficiency through the optical range.
  
  Figure~\ref{fig_ddc15_tau_at_vel} does not show the opacity associated with the stripped material itself. For model DDC15 with $M_{\rm st}=$\,0.18\,\msun, the Rosseland-mean (electron-scattering) optical depth at the base of the ejecta (considering the whole column of material above the innermost ejecta layer at 200\,\kms) drops from 3.4 to 0.22 (from 2.54 to 0.17) between 100 and 300\,d.  This is a greater reduction than expected for homologous expansion (wherein $\tau \propto 1/t^2$) because of a simultaneous reduction in ionization. This dependence on ionization level usually applies when electron scattering dominates, but in some cases opacity can rise with decreasing ionization if this shift leads to a greater supply of optically thick lines.

  An additional ingredient is the optical depth of the line itself (its intrinsic optical depth). The stripped material is quite abundant (representing up to a third of the SN Ia ejecta mass, for example, if we consider $M_{\rm st}$ of $0.5$\,\msun) and is also located at the lowest velocities, hence the highest densities. For model DDC15 with $M_{\rm st}=$\,0.18\,\msun, we find that H$\alpha$ has an optical depth of 100 or more in its formation region over the period $100-300$\,d. The H$\alpha$ optical depth becomes small in regions where the H mass fraction is small or negligible (thus contributing no emission). Hence, H$\alpha$ is optically thick at all times (even if $M_{\rm st}$ is reduced by a factor of 100). The same holds but to a smaller degree for H$\beta$. This implies that optically thin line formation for Balmer lines is not a valid assumption. This may be one origin for the differences in the results of \citet{mattila_01el_ha_05} and those of \citet{botyanszki_ia_neb_sd}.

  To conclude, our simulations of stripped material in SN Ia ejecta suggest that the H$\alpha$ line (and emission from the stripped material) suffers from a brightness contrast with the metal-rich ejecta as well as optical depth effects that are both intrinsic and extrinsic. As a consequence, the H$\alpha$ line, which is the strongest signature from the stripped material, is essentially impossible to detect prior to 50\,d (metal-rich ejecta are optically thick and too luminous), hard to detect at 100\,d (metal-rich ejecta are not as optically thick but still very luminous) but increasingly strong as time progresses (metal-rich ejecta become increasingly transparent and faint). The easiest line to detect is H$\alpha$ in part because it lies in a region largely devoid of strong or optically thick metal lines.

\begin{figure*}
\begin{center}
\includegraphics[width=0.8\hsize]{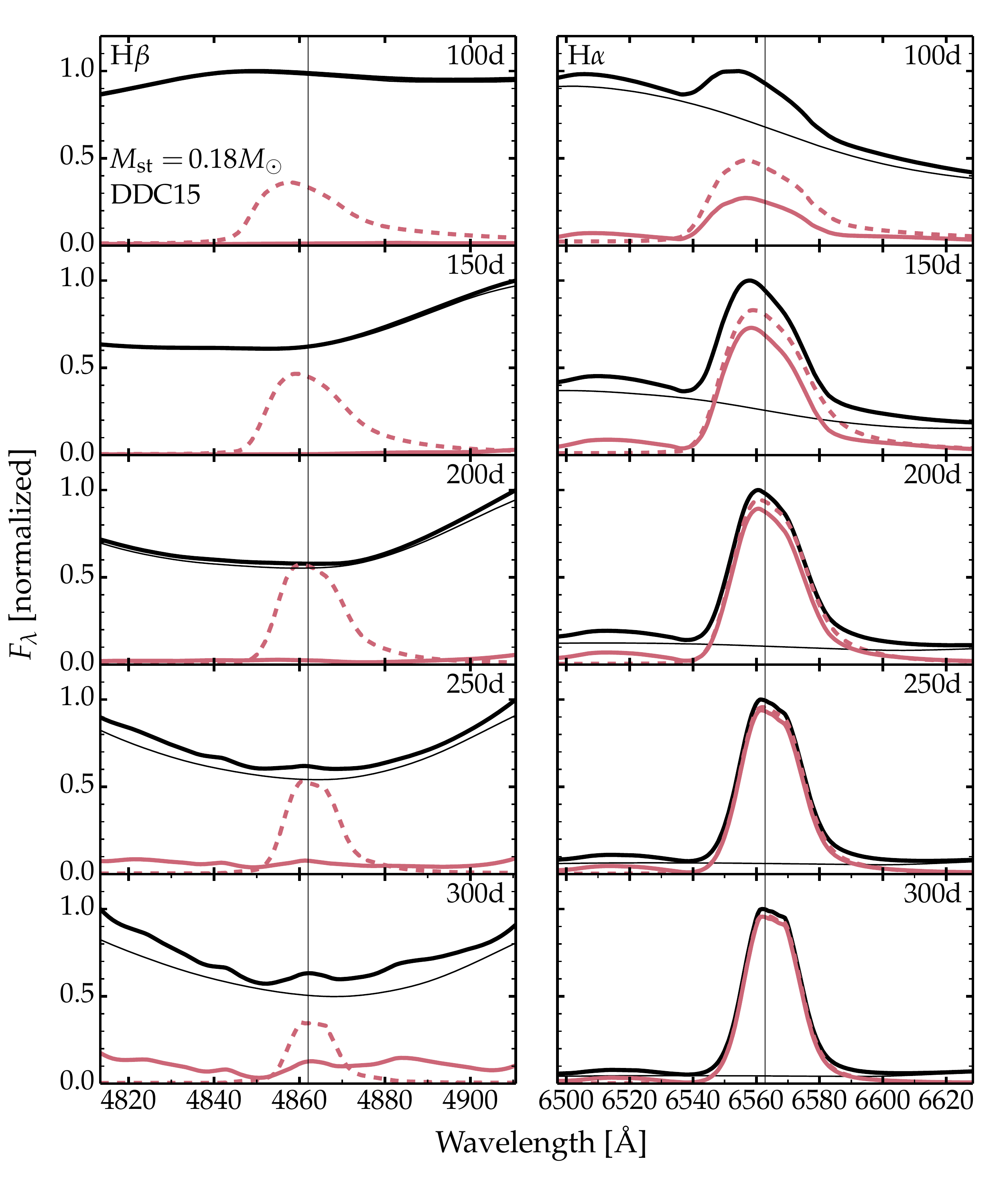}
\end{center}
\vspace{-1cm}
\caption{Comparison of the spectral regions around H$\beta$ (left) and H$\alpha$ (right) for model DDC15 with $M_{\rm st}=$\,0.18\,\msun\ (thick black curve) and $M_{\rm st}=$\,10$^{-5}$\,\msun\ (thin black curve; the stripped material contributes negligibly in that case). The different epochs spanning 100 to 300\,d are stacked vertically. The difference between the two models gives a physical measure of the contribution from the stripped material (thick red line). For comparison, we show the H\one\ spectrum that results if we neglect the influence of all other ions (thick dashed red line). It can thus be used to gauge the influence of the metal-rich ejecta on the emission from the stripped material. This influence is moderate or negligible for H$\alpha$, but very strong for H$\beta$. This figure shows that stripped material is not expected to yield any observable H$\beta$ emission for hundreds of days because of metal-line blanketing from the metal-rich (SN Ia) ejecta, whereas H$\alpha$ emission is prominent.
\label{fig_gun}
}
\end{figure*}

Optical depth effects also cause a blueshift of line profiles, which is most obviously seen through the location of peak flux in the line. This effect is well understood \citep{DH05a} and routinely observed in a variety of SNe \citep{blondin_snia_06,anderson_blueshift_14,zhang_10jl_12}. Here, the effect is present over the period $100-200$\,d after explosion (see Fig.~\ref{fig_muti_panel_Ha_DDC15} and the next section) for cases in which $M_{\rm st}$ is greater than about 0.1\,\msun. Hence, line shifts may not result exclusively from an asymmetric ejecta. This complicates the identification of multidimensional effects \citep{botyanszki_ia_neb_sd}.

\section{Conclusive evidence for the identification of stripped material}
\label{sect_gun}

A concern from the observations of H$\alpha$ in SNe Ia at $100-300$\,d after explosion is whether the emission arises from the presence of stripped material in the innermost ejecta layers (the context explored here) or whether the emission is external to the ejecta and arises from interaction with the H-rich CSM (produced by wind mass loss and mass transfer in the progenitor system). The results described above provide some clues to this problem.

Figure~\ref{fig_gun} presents the evolution of the H$\beta$ and H$\alpha$ regions at 100, 150, 200, 250, and 300\,d after explosion in the model DDC15 with $M_{\rm st}=$\,0.18\,\msun. The total flux (thick black curve) is shown together with the predictions of the model counterpart with $M_{\rm st}=$\,10$^{-5}$\,\msun\ (thin black curve). If we subtract the two, we can see the contribution from H$\beta$ and H$\alpha$ as a function of time in the model with 0.18\,\msun\ of stripped material (thick red line). The contribution to H$\alpha$ is significant at all times, and represents the total strength of the feature at 6562\,\AA\ at 300\,d since it coincides with the total flux. For H$\beta$, the line is absent until 200\,d and only rises as a very weak feature at $250-300$\,d. The presence of H$\alpha$ and the absence of H$\beta$ is caused by metal-line blanketing. If we compute the H\one\ spectrum by ignoring all other ions (and associated emission and absorption), we see that both H$\alpha$ and H$\beta$ are present (thick red dashed line). In each case, the line also peaks blueward of its rest wavelength up until 200\,d, indicating that it forms under optically thick conditions.

The strong metal-line blanketing by the metal-rich ejecta thus offers a clear indication of the presence of stripped material since in that case only H$\alpha$ should be seen. H$\beta$ photons are destroyed by metal lines at all times prior to about 300\,d and should not be seen if the emission arises from stripped material at low velocity. Hence, the detection of H$\beta$ in SNe Ia suggests that the emission arises from regions outside the SN Ia ejecta, free of metal-line blanketing, and thus most likely arising from interaction with CSM. High signal to noise ratio ($S/N)$ observations of the optical and in particular the H$\beta$ region are essential to distinguish both scenarios.

\section{Influence of the adopted stripped material velocity and density}
\label{sect_offset}

In our simulations, we assumed spherical symmetry and adopted a centrally concentrated distribution for the stripped material. The outer bound of this region is around 1000\,\kms. Multidimensional hydrodynamical simulations suggest that the stripped material can be offset to larger velocities, as well as offset from the ejecta center \citep{marietta_snia_hyd_00,pakmor_snia_ms_08,liu_snia_12,pan_snia_12,botyanszki_ia_neb_sd}. To test this effect, we recomputed model DDC15 with $M_{\rm st}=$\,0.18\,\msun\ but using a distribution of stripped material centered around (rather than bounded by) 1000\,\kms, extending from 200\,\kms\ up to about 2000\,\kms. The stripped material is now a hollow shell, occupying a larger volume, but  still symmetric around the ejecta center. The gaussian smoothing does not conserve the mass exactly in the innermost layers (because of the density profile at the inner boundary) so the stripped material mass is 30\% lower in the ``offset" model.

\begin{figure*}
\includegraphics[width=\hsize]{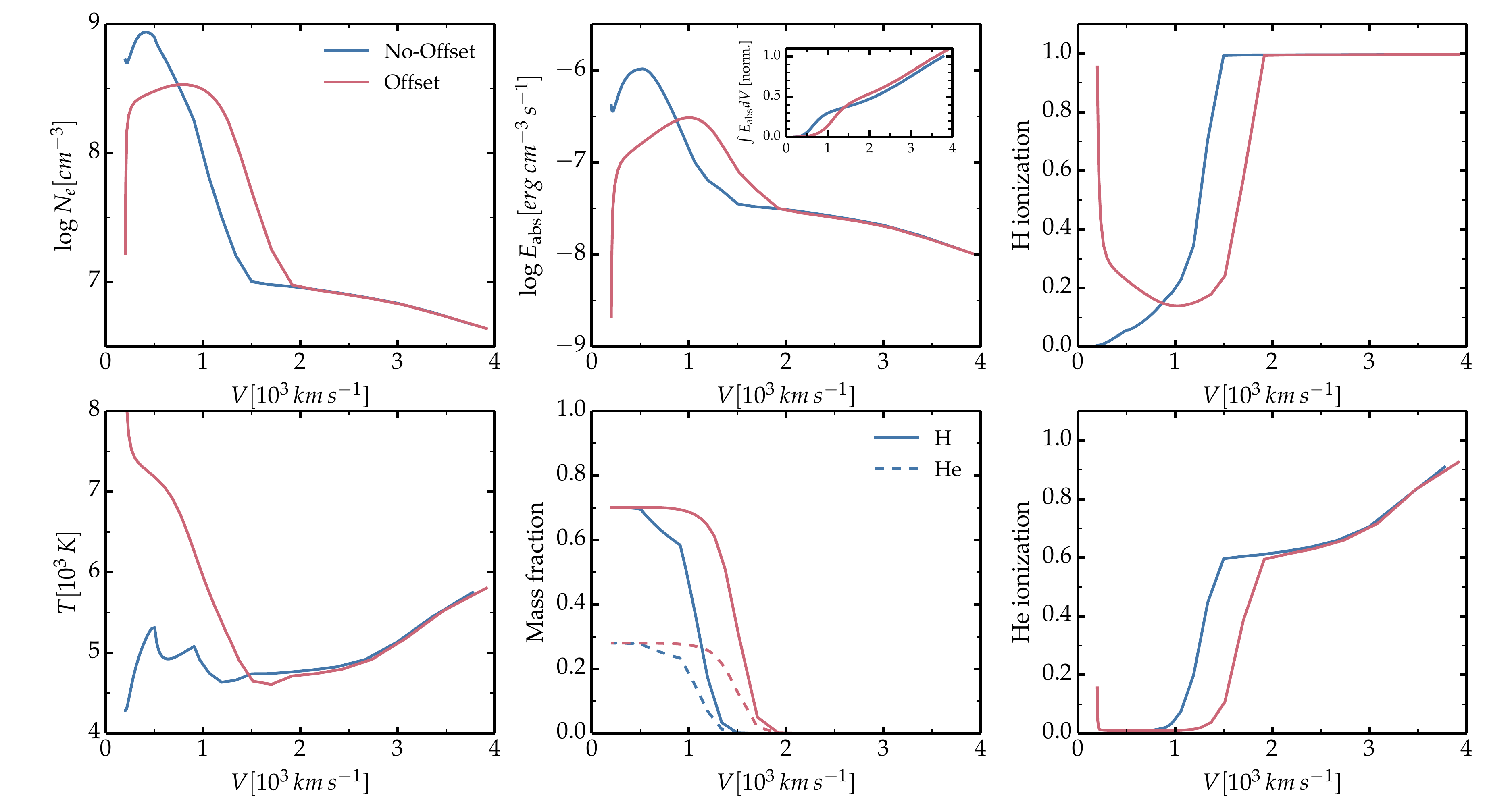}
\caption{Impact of the adopted density profile for the stripped material on the gas properties in model DDC15. This model has a $M_{\rm st}$ of about 0.15\,\msun\ (the value is 30\% lower in the model with offset) and at a SN age of 200\,d. The simulations are 1D (i.e., assume spherical symmetry), but the simulation ``No-offset" uses a more centrally concentrated mass distribution than the simulation ``Offset'' (see the electron-density profile, which reflects closely the difference in mass density profile). In the inset of the top-middle panel, the normalization of the cumulative decay power absorbed is set to the value obtained for the model with no offset (and limited to the velocity region shown).
\label{fig_offset_ejecta}
}
\end{figure*}

Figure~\ref{fig_offset_ejecta} shows the impact of the stripped material distribution on the ejecta properties. With the offset, the stripped material is shifted to larger velocities, corresponding to lower densities (aggravated by the 30\% lower stripped material mass). The electron density shows a broader bump for this H-rich material, with a slightly higher temperature, and a higher mean ionization for H (same ionization for He). The decay power absorbed by the stripped material is comparable in both models.

\begin{figure}
\includegraphics[width=\hsize]{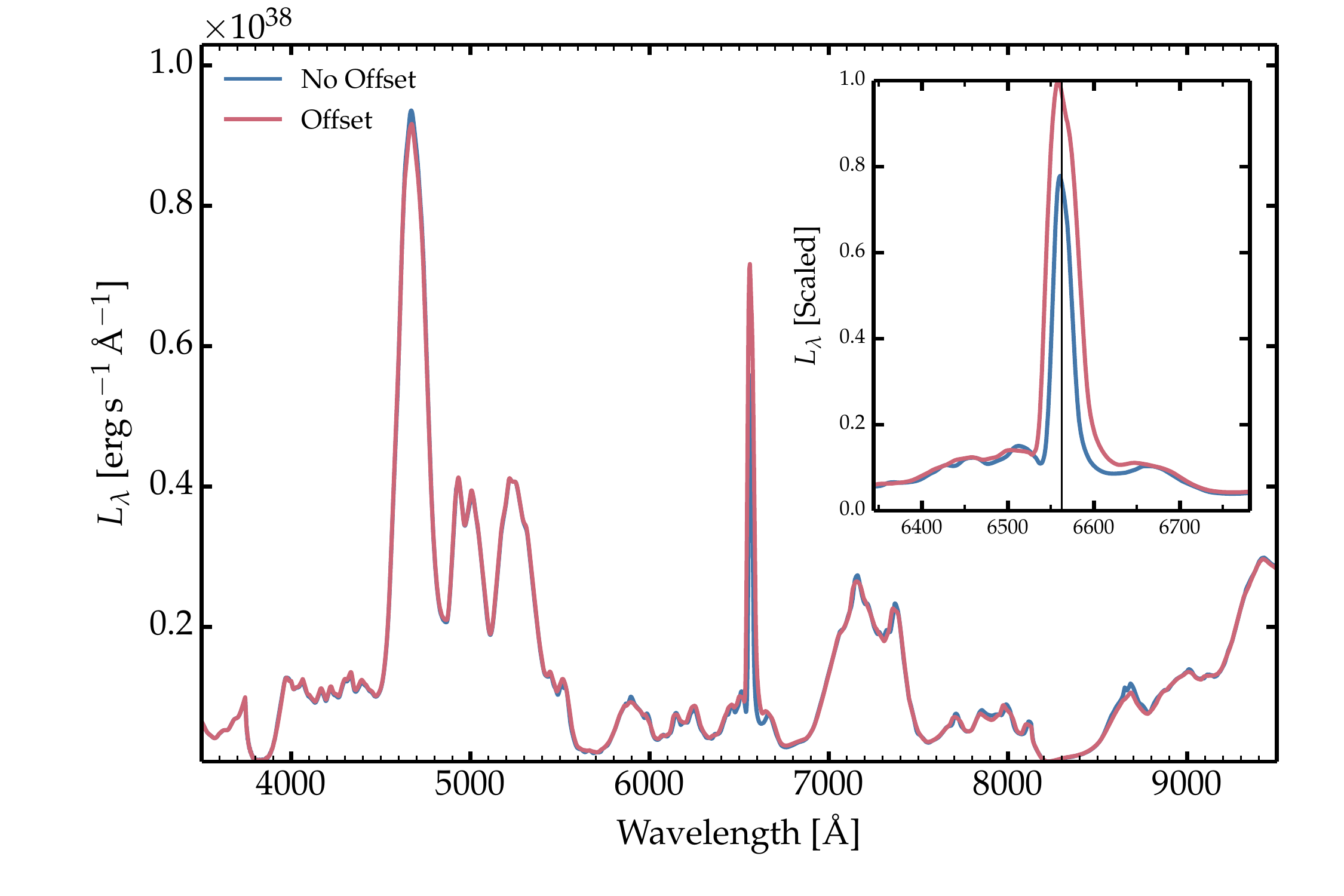}
\vspace{-0.5cm}
\caption{Same as Fig.~\ref{fig_offset_ejecta}, but now showing the impact of the adopted density profile for the stripped material  on the optical radiation.
\label{fig_offset_spec}
}
\end{figure}

However, the H$\alpha$ line in the ``offset" model is twice as strong in total line flux and  30\% stronger in peak flux (Fig.~\ref{fig_offset_spec}). The H$\alpha$ line is also broader because of its formation at higher velocities. The change in strength likely arises from a formation at lower optical depth, both in the continuum and in the line. The formation over a larger volume probably contributes to desaturate the line since H$\alpha$ photons are emitted over a broader range of velocities.

The choice of parametrized configurations that we use here, in the spirit of \citet{mattila_01el_ha_05}, can itself lead to significant variations  in H$\alpha$ line strength, so this needs to be borne in mind when estimating the stripped material mass from observations. The factor of two found here, however, is  much smaller than the difference between predicted values for $M_{\rm st}$ in the single-degenerate scenario and the inferred values for $M_{\rm st}$ from observations.

\begin{figure*}
\begin{center}
\includegraphics[width=0.8\hsize]{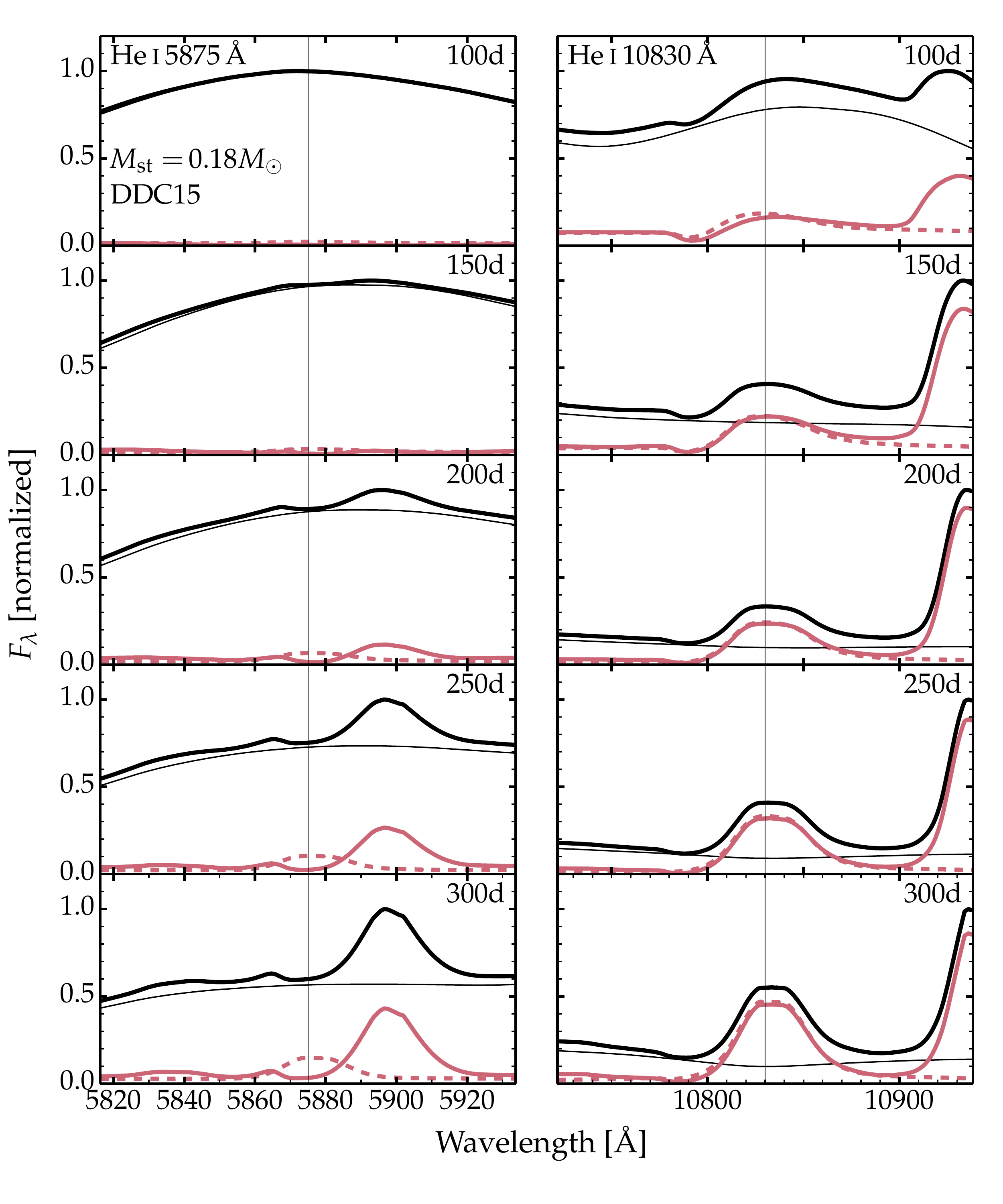}
\end{center}
\vspace{-1cm}
\caption{Same as Fig.~\ref{fig_gun}, but for the spectral regions centered on He\one\,5875\,\AA\ (left) and He\one\,10830\,\AA\ (right). He\one\,5875\,\AA\ is absent in the total emergent spectrum, and at best weak at late times in the He\one-only spectrum. The only weak feature we predict is Na\one\,D. He\one\,10830\,\AA\ is present at all times, is strong, and largely unaffected by the metal-rich ejecta (in either emission or absorption).
\label{fig_he1}
}
\end{figure*}

\section{He\one\ lines}
\label{sect_he1}

He\one\ lines are non-thermally excited in our simulations and are predicted to be present but weak in the He\one-only spectra that we compute (Fig.~\ref{fig_indiv_spec}). In the total emergent spectrum, the weak feature at 5900\,\AA\ is due to Na\one\,D, while He\one\,5875\,\AA\ is absent (left panel of Fig.~\ref{fig_he1}). The only unambiguous He signature is He\one\,10830\,\AA\ in the near infrared (right panel of Fig.~\ref{fig_he1}) because this line is usually the strongest under similar SN ejecta conditions \citep{li_etal_12_nonte}. Metal-line blanketing is also weaker in the near infrared and the stripped material is only a few times fainter than the metal-rich ejecta in this spectral region.  The He\one\,10830\,\AA\ line is not affected by metal-line blanketing even at 100\,d after explosion (this holds because the solid and dashed lines essentially overlap; see Fig.~\ref{fig_gun} for explanations).

This result is not surprising since the properties of the stripped material in a SN Ia ejecta at such late times are analogous to those found in Type II SN ejecta at the same epoch. In general, He\one\ lines are not observed or predicted in optical spectra of type II SNe at nebular times (see, for example, \citealt{silverman_neb_17}, \citealt{jerkstrand_04et_12}; the models of \citet{d13_sn2p} predict the presence of He\one\,7065\,\AA\  and overestimate its strength, but the problem is caused by the adoption of a high turbulent velocity not used here; Dessart \& Hillier in prep.).

Our results disagree with the predictions of \citet{botyanszki_ia_neb_sd}. This may arise from their neglect of optical depth effects (especially in the optical). We also find that a significant power absorbed by the stripped material emerges in the continuum (or in lines that are later reprocessed by the metal rich ejecta), while they seem to assume that the gas can only cool through (optically thin) line emission. 

\begin{figure}
\includegraphics[width=\hsize]{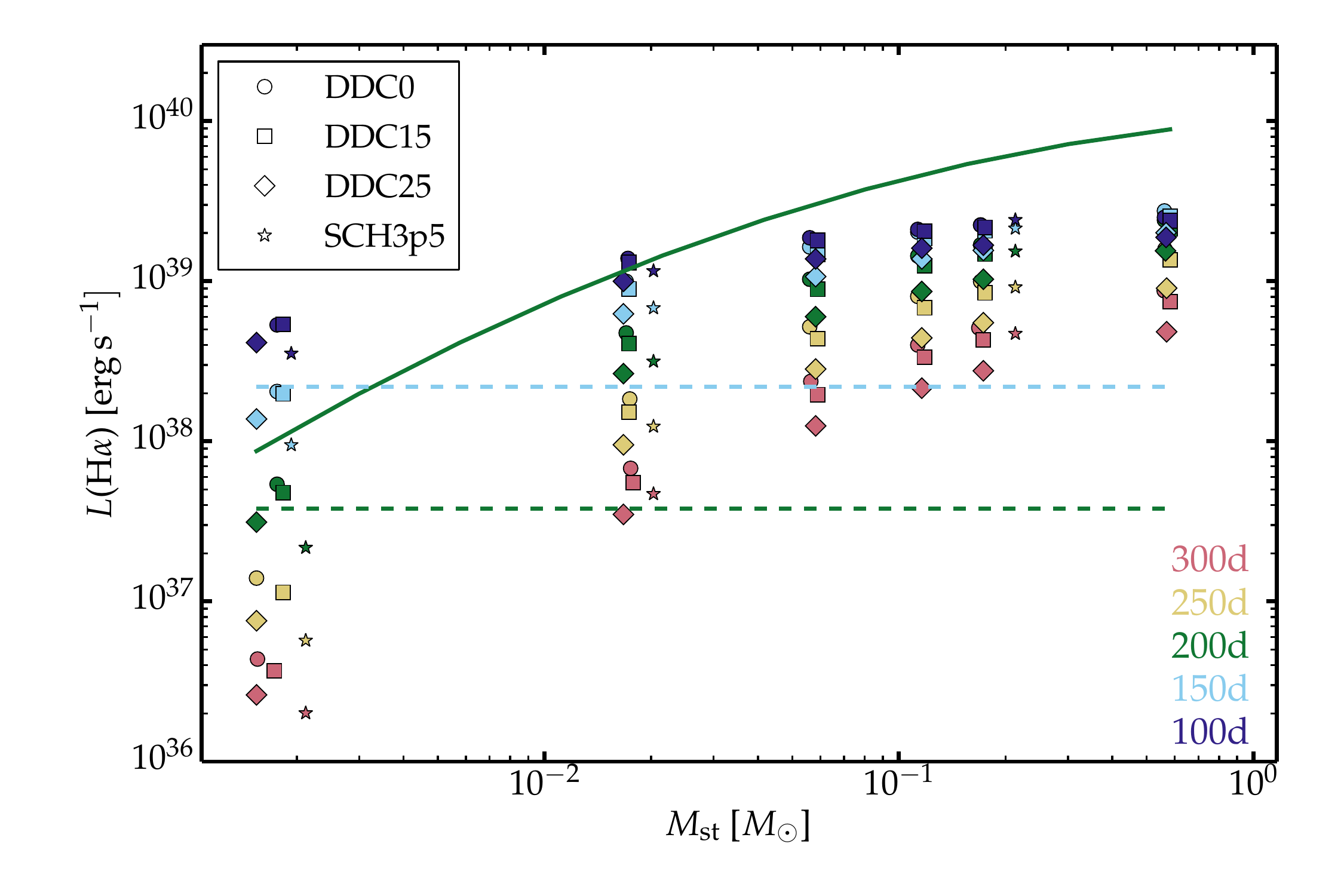}
\includegraphics[width=\hsize]{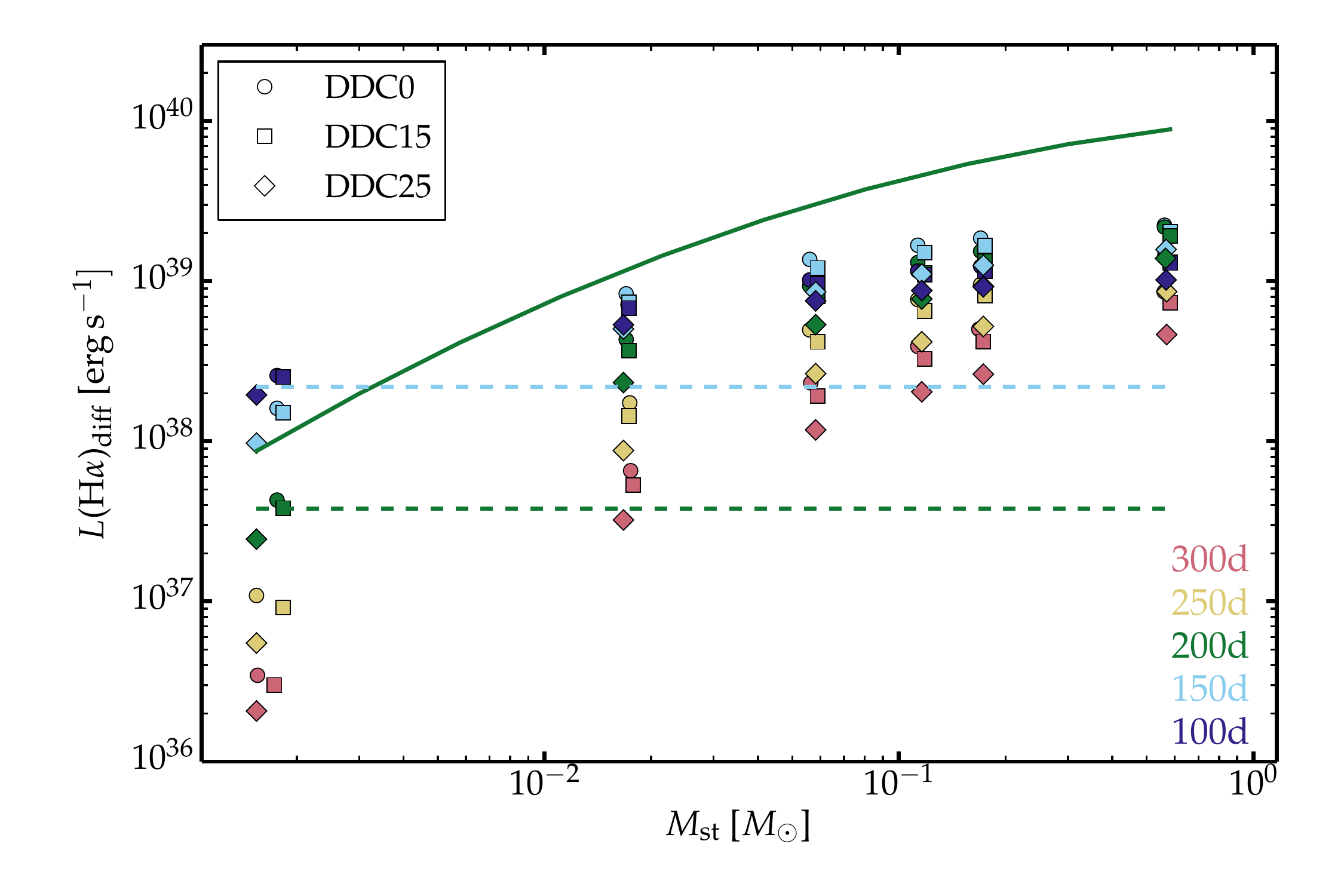}
\vspace{-0.5cm}
\caption{Top: Variation of the H$\alpha$ luminosity as a function of the adopted stripped-material mass $M_{\rm st}$ for our set of $M_{\rm ch}$ delayed-detonation simulations DDC0 (\nifs\ mass is 0.86\,\msun), DDC15 (\nifs\ mass is 0.51\,\msun), DDC25 (\nifs\ mass is 0.12\,\msun), and sub-$M_{\rm ch}$ detonation model SCH3p5 (\nifs\ mass is 0.30\,\msun). The calculation is based on the H\one\ spectrum computed by neglecting all other ions and thus ignores blanketing effects from the metal-rich ejecta. The color coding defines the SN age, from 100 to 300\,d after explosion. The solid (green) curve corresponds to the calibration of \citet{botyanszki_ia_neb_sd} based on their model calculations (and corrected for the typo in their Eq.~1). The dashed lines correspond to the observed H$\alpha$ luminosity in ASASSN-18tb   at $\sim$\,153\,d (\citealt{kollmeier_18tb_ha_19}; dashed line drawn in turquoise to match the color used for models computed at 150\,d) and SN\,2018cqj  at $\sim$\,207\,d (\citealt{prieto_var_ha_19}; dashed line drawn in green to match the color used for models computed at 200\,d), compatible with our models that have a stripped-material mass of about 0.002\,\msun. Bottom: Same as top (with the omission of results for SCH3p5 models), but we compute the H$\alpha$ luminosity as given by subtracting the model with no stripped material (i.e., the model with $M_{\rm st}=$\,10$^{-5}$\,\msun) from the model with the stripped material. This yields a lower H$\alpha$ luminosity by up to a factor of two at times prior to 200\,d, but makes little difference later on (see Fig.~\ref{fig_gun} and discussion in Sect.~\ref{sect_gun}).
\label{fig_mh_LHa}
}
\end{figure}

\section{Results for the grid of models}
\label{sect_grid}

We now turn to the presentation of results for the whole grid of simulations. We have performed radiative transfer calculations at 100, 150, 200, 250, and 300\,d for models DDC0 (\nifs\ mass is 0.86\,\msun), DDC15 (\nifs\ mass is 0.51\,\msun), DDC25 (\nifs\ mass is 0.12\,\msun), and SCH3p5 (\nifs\ mass is 0.30\,\msun) and stripped material masses of 0.0018, 0.018, and 0.18\,\msun. For each SN Ia model and epoch, we also computed the cases of a very low stripped material mass of 10$^{-5}$\,\msun\ to facilitate the assessment of the contribution of the stripped material when its influence is weak. For models DDC0, DDC15, and DDC25, additional simulations were also done to better cover the region around $M_{\rm st}=$\,0.1\,\msun. In total, the grid comprises more than a hundred simulations and encompasses a much wider range of values in \nifs\ mass and  $M_{\rm st}$ than studied so far.

\subsection{H$\alpha$ luminosity and comparison to observations}

Figure~\ref{fig_mh_LHa}  shows the H$\alpha$ luminosity versus stripped material mass for all SN Ia ejecta models (symbols) and epochs (color coding); polynomial fits to the model results are provided in Appendix~\ref{appendix_fit}. Two different ways are used to compute the H$\alpha$ luminosity. For the top panel, we computed the H\one\ spectrum and integrated over H$\alpha$ to obtain the corresponding line flux and luminosity. For the bottom panel, we computed the total emergent spectrum and subtracted the spectrum for the model counterpart with $M_{\rm st}=$\,10$^{-5}$\,\msun.  The two methods differ by up to a factor of two at times prior to 200\,d (Sect.~\ref{sect_gun} and Fig.~\ref{fig_gun}).

Computed either way, the H$\alpha$ luminosity reaches a maximum of a few 10$^{39}$\,\ergs\ at 100\,d for the largest values of $M_{\rm st}$. This luminosity tends to drop as time progresses (although in a complicated way; see below) or with stripped material mass. Variations in \nifs\ mass between models lead to similar variations in H$\alpha$ luminosity. The general pattern shown in log-log space does not allow a fine assessment. However, it is clear that the inferred H$\alpha$ line luminosities by \citet{kollmeier_18tb_ha_19} and \citet{prieto_var_ha_19} for ASASSN-18tb and SN\,2018cqj are both compatible with our models with $M_{\rm st}$ of about 0.002\,\msun, which is in agreement with their analysis.

Our H$\alpha$ luminosities are lower by a factor of five to ten than those of \citet{botyanszki_ia_neb_sd}, which are given at a SN age of 200\,d (green solid line in Fig.~\ref{fig_mh_LHa}). This offset probably stems in part from the differences in the hydrodynamical structure of the ejecta, since their computed ejecta configurations tend to have the stripped material offset at a higher velocity than adopted in our work (see results and discussion in Sect.~\ref{sect_offset}). It is likely that differences also result from optical depth effects (Sect.~\ref{sect_ha_appearance}), or differences in the model atom and metal composition for the stripped material (metals can play an important part in the cooling of the gas even at low abundance). 

\subsection{H$\alpha$ luminosity versus decay power absorbed}
\label{sect_eabs_ha}

\begin{figure}
\includegraphics[width=\hsize]{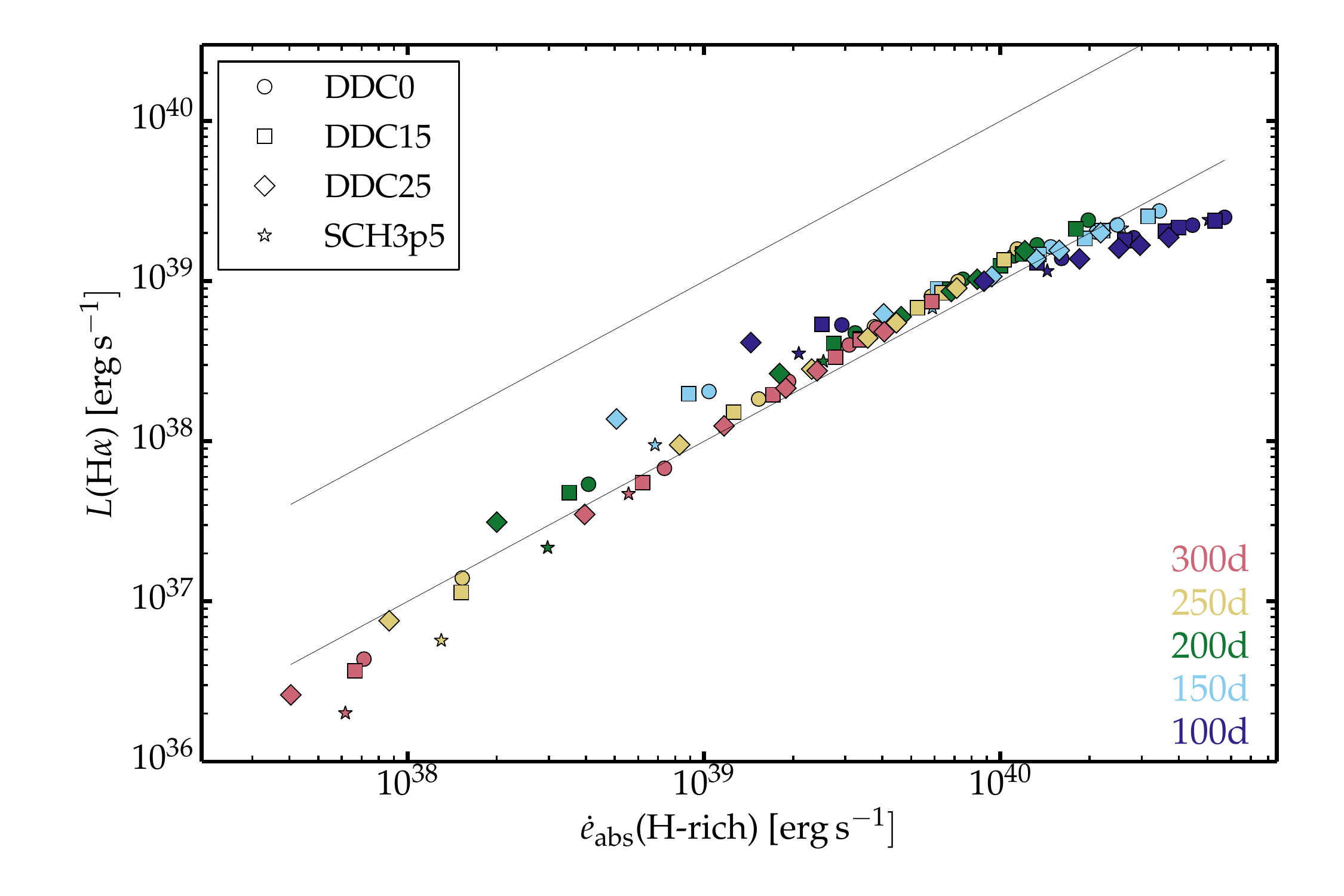}
\vspace{-0.5cm}
\caption{H$\alpha$ luminosity $L$(H$\alpha$) vs. decay power absorbed by the H-rich stripped material from the companion. The two black lines correspond to the cases where $L(H\alpha) = \dot{e}_{\rm abs}$(H-rich) and $L(H\alpha) = 0.1\dot{e}_{\rm abs}$(H-rich). For most models, H$\alpha$ radiates about 10\% (the distribution covers from 5\%\ to 30\% due to several outliers, all for a low stripped material mass) of the decay power absorbed by the stripped material (this material is \nifs\ deficient and thus the power stems from non-local $\gamma$-ray energy deposition). 
\label{fig_eabs_h_LHa}
}
\end{figure}

\begin{figure}
\includegraphics[width=\hsize]{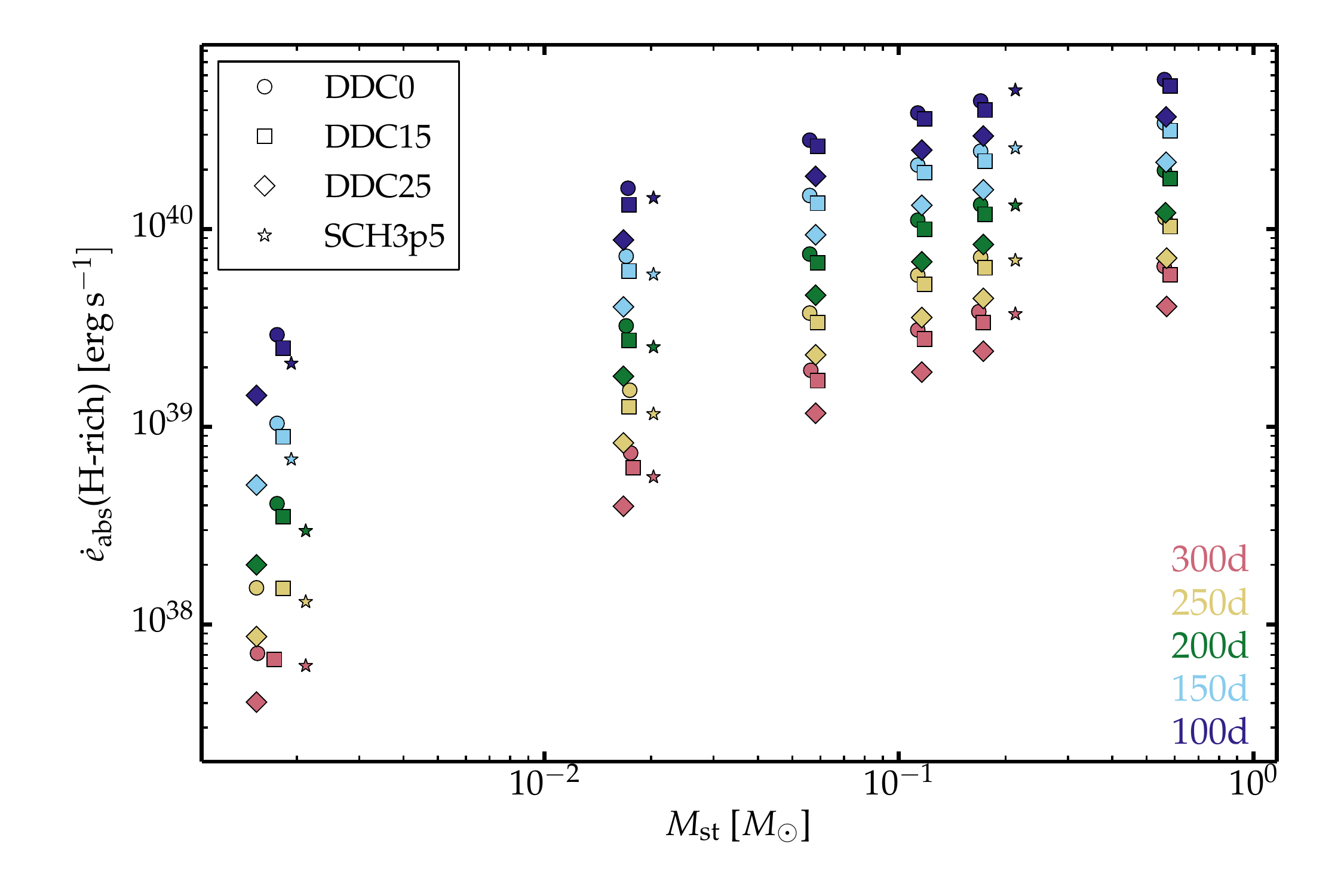}
\caption{Decay power absorbed by the stripped material vs. its mass for our set of models. As before, the color coding distinguishes the SN ages, which cover from 100 to 300\,d. The decay power absorbed is greater at earlier times (because the total decay power is greater earlier) or for greater masses of stripped material (because of the greater trapping). However, this does not imply that the H$\alpha$ line is more easily seen earlier since the SN luminosity is also greater. The line luminosity may be high but the equivalent width, or pseudo-equivalent width, may be small and even null.
\label{fig_mh_eabs_h}
}
\end{figure}

We find that the H$\alpha$ line radiates about 10\% (extending from 5\%\ up to 30\% for a few outliers) of the decay power absorbed by the stripped material (Fig.~\ref{fig_eabs_h_LHa}), with no clear dependence with SN age, $M_{\rm st}$, or \nifs\ mass. In particular, this fraction stays about constant despite the large variations in decay power absorbed with SN age and with  $M_{\rm st}$ (Fig.~\ref{fig_mh_eabs_h}).

Here, we quote the H$\alpha$ flux from an H\one\ spectrum calculation, thus without any influence from the metal-rich ejecta. By doing this, we can gauge the cooling power of H$\alpha$ for the stripped material. This implies that 90\% of the decay power absorbed by the stripped material is radiated by other means. A fraction of this power goes in lines, and the rest in continuum radiation (see Fig.~\ref{fig_indiv_spec} for one representative case). Alternatively, we could quote the H$\alpha$ luminosity from the total emergent spectrum. This is more useful to compare to observations, but prevents a proper evaluation of the cooling power of H$\alpha$ since the H$\alpha$ emission from the stripped material is reprocessed by the metal-rich ejecta.

The fraction $L$(H$\alpha$)/$\dot{e}_{\rm abs}$(H-rich) of about 10\% holds over three orders of magnitude in decay power absorbed by the stripped material. This is lower than the value of 30\% found by \citet{botyanszki_ia_neb_sd} for most of their models. 

\begin{figure}[ht!]
\includegraphics[width=\hsize]{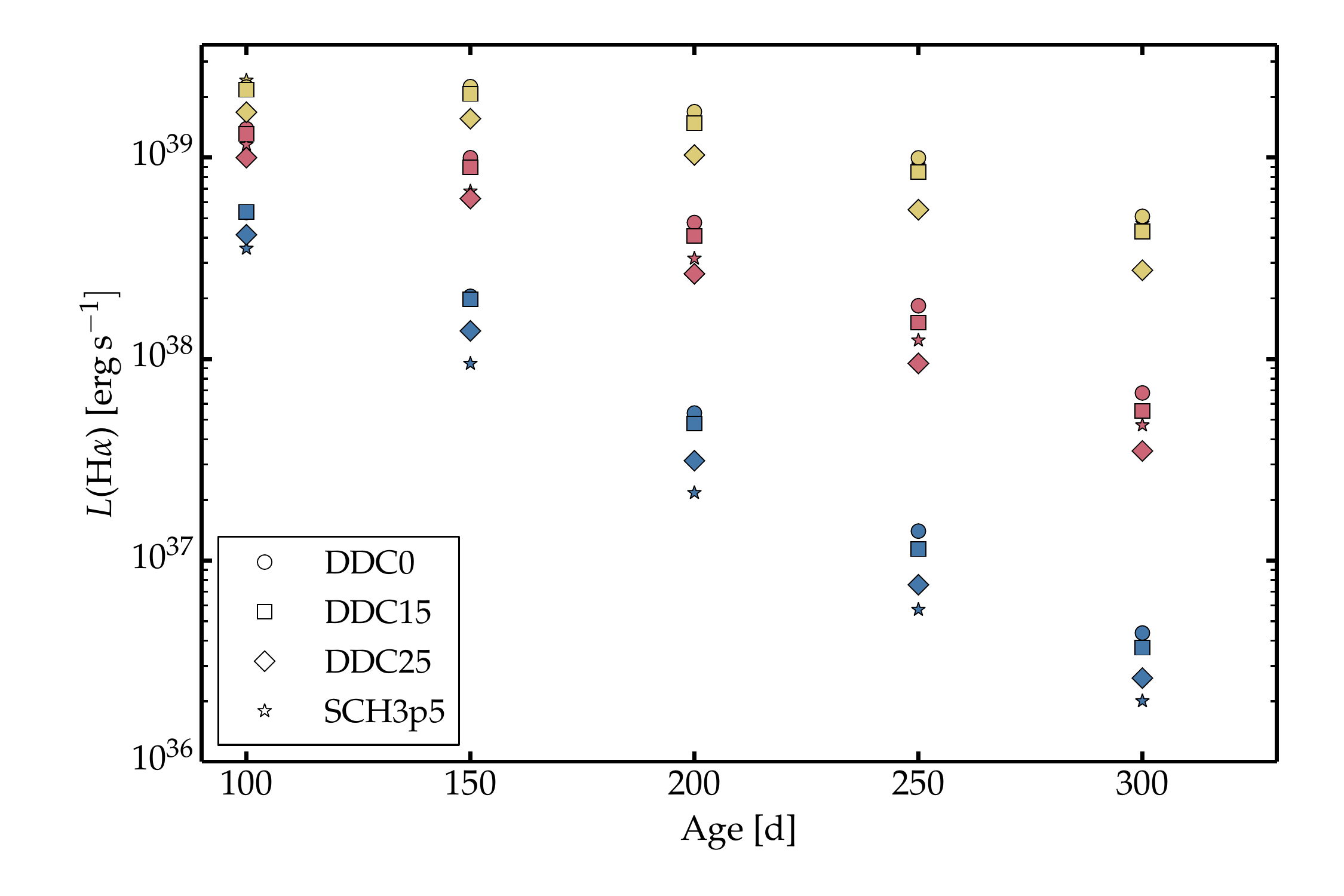}
\includegraphics[width=\hsize]{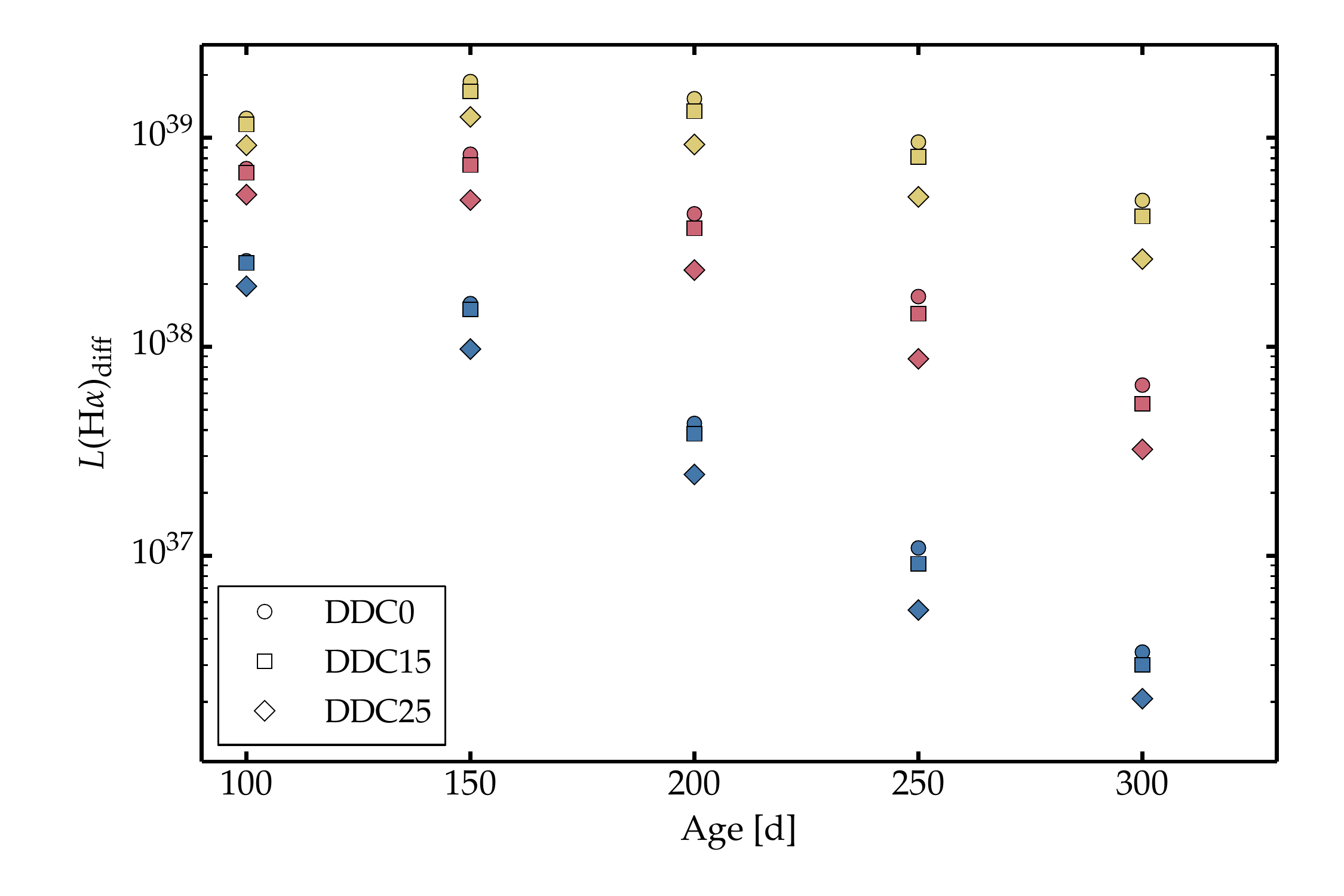}
\includegraphics[width=\hsize]{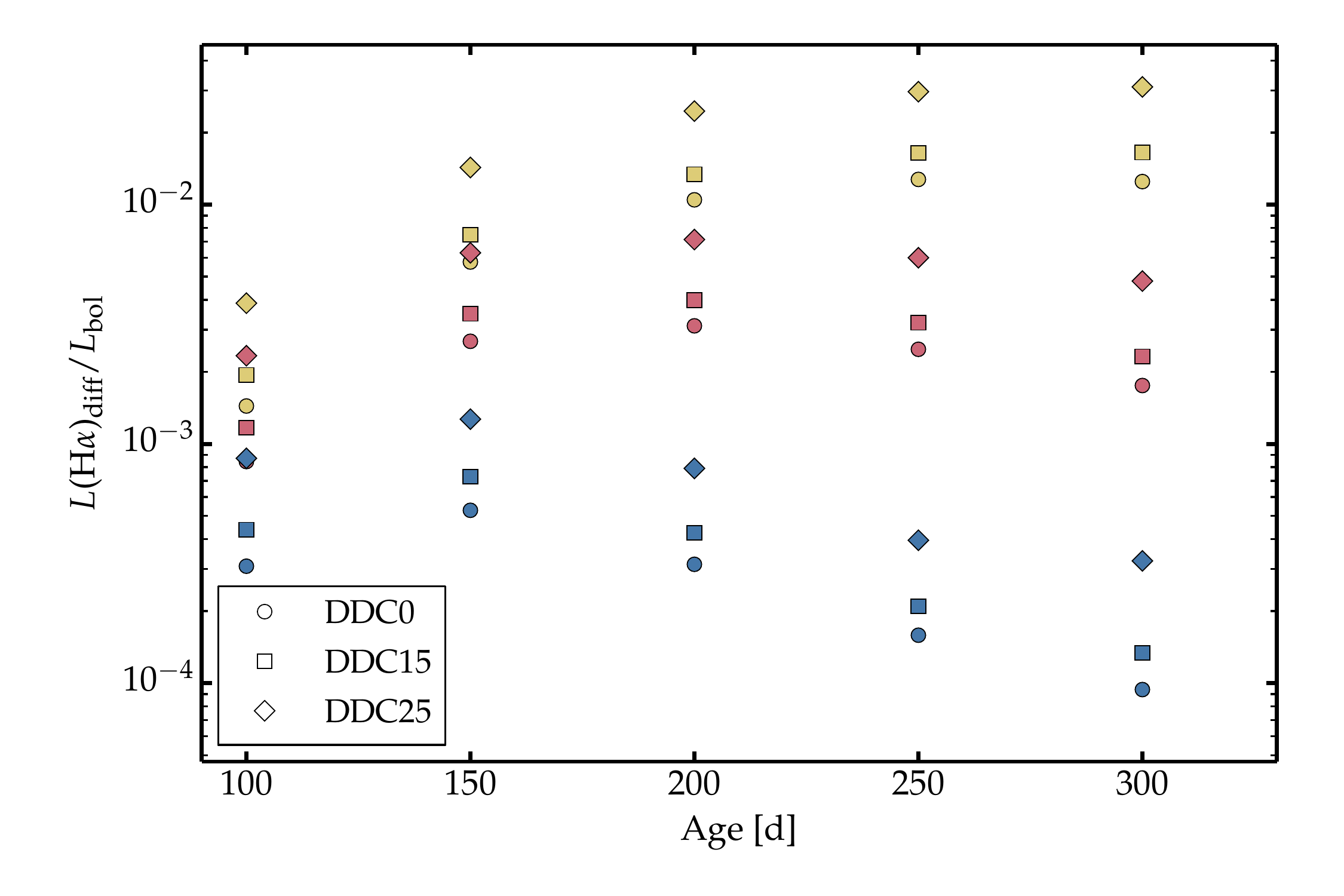}
\caption{Evolution with SN age of the H$\alpha$ luminosity (top and middle) or the normalized H$\alpha$ luminosity (bottom) for a subset of models covering stripped material masses of about 0.0018 (blue), 0.018 (red), and 0.18\,\msun\ (yellow). For the two lower panels, the H$\alpha$ flux is calculated by integrating over H$\alpha$ after subtracting the flux from the corresponding reference model with $M_{\rm st}=$\,10$^{-5}$\,\msun. For the top panel, the H$\alpha$ flux is calculated from the H\one\ spectrum. The different behavior with stripped material masses arises from optical depth effects, which are important in the time span 100 to 200\,d here. These effects can produce an H$\alpha$ luminosity that evolves very differently from that expected for \cofs\ decay, even though \cofs\ decay is the power source for the emission from the stripped material.
\label{fig_age_LHa}
}
\end{figure}

\subsection{Evolution of the H$\alpha$ luminosity}
\label{sect_ha_evol}

We have seen that the H$\alpha$ luminosity represents about 10\% of the decay power absorbed (Fig.~\ref{fig_eabs_h_LHa}). This dependence implies a connection to the \cofs\ characteristic decay time. However, there are a number of complications that can yield an H$\alpha$ luminosity that deviates from the evolution of the decay power emitted or absorbed. These complications can be, for example, a different luminosity evolution for the metal-rich ejecta and for the stripped material, combined with the evolution of the blanketing effect from the metal-rich ejecta.

The top and middle  panels of Fig.~\ref{fig_age_LHa} show the evolution of the H$\alpha$ luminosity for $M_{\rm st}$ of 0.0018, 0.018, and 0.18\,\msun, first based on the H\one\ spectrum and then based on a difference between the model counterparts with $M_{\rm st}=$\,10$^{-5}$\msun. This evolution is smooth and qualitatively similar for all cases, largely irrespective of \nifs\ mass. However, the models with lower $M_{\rm st}$ decline faster, along a nearly constant slope (in the log, so in magnitude). The models with a higher $M_{\rm st}$ show a plateau from 100 to 200\,d (or even a rise if the flux difference is used), and then decline, but more slowly.  The different behavior seen at early times in both panels arises from optical depth effects. It shows that up to 200\,d, a very slow decline in the H$\alpha$ luminosity can arise even if the power source is \cofs\ decay; it does not necessarily imply CSM interaction, as proposed by \citet{vallely_snian_19}.

When normalized to the bolometric luminosity (which is equal to the total decay power absorbed at these late epochs), the H$\alpha$ luminosity is constant or growing for models with a high $M_{\rm st}$ but decreases with time otherwise. This is in part related to the fact that at low $M_{\rm st}$, $\gamma$-rays are not efficiently trapped by the stripped material, while the SN Ia ejecta are increasingly powered by local positron energy (which the stripped material cannot receive since it is \nifs\ deficient). A corollary is that the stripped material allows the SN to be more luminous since it contributes to trapping $\gamma$-rays that would have otherwise escaped. This is not necessarily seen in H$\alpha$, but may yield a continuum flux excess throughout the optical. This is visible in the bottom-row panels of Fig.~\ref{fig_muti_panel_Ha_DDC15} where the models with higher $M_{\rm st}$ are offset to higher luminosities at all wavelengths, not just in strong lines like H$\alpha$.

In the upper panel of Fig.~\ref{fig_age_LHa}, we computed the line luminosity from the H\one\ spectrum in order to understand how this luminosity evolves without the corrupting effect of photon reprocessing by the metal-rich ejecta. However, this reprocessing is moderate after about 200\,d for H$\alpha$. After that time, the results shown in Fig.~\ref{fig_age_LHa} reflect closely the evolution of the emergent H$\alpha$ luminosity (see Fig.~\ref{fig_gun}).

\subsection{H$\alpha$ versus H$\beta$ luminosities}
\label{sect_ha_hb}

\begin{figure}
\includegraphics[width=\hsize]{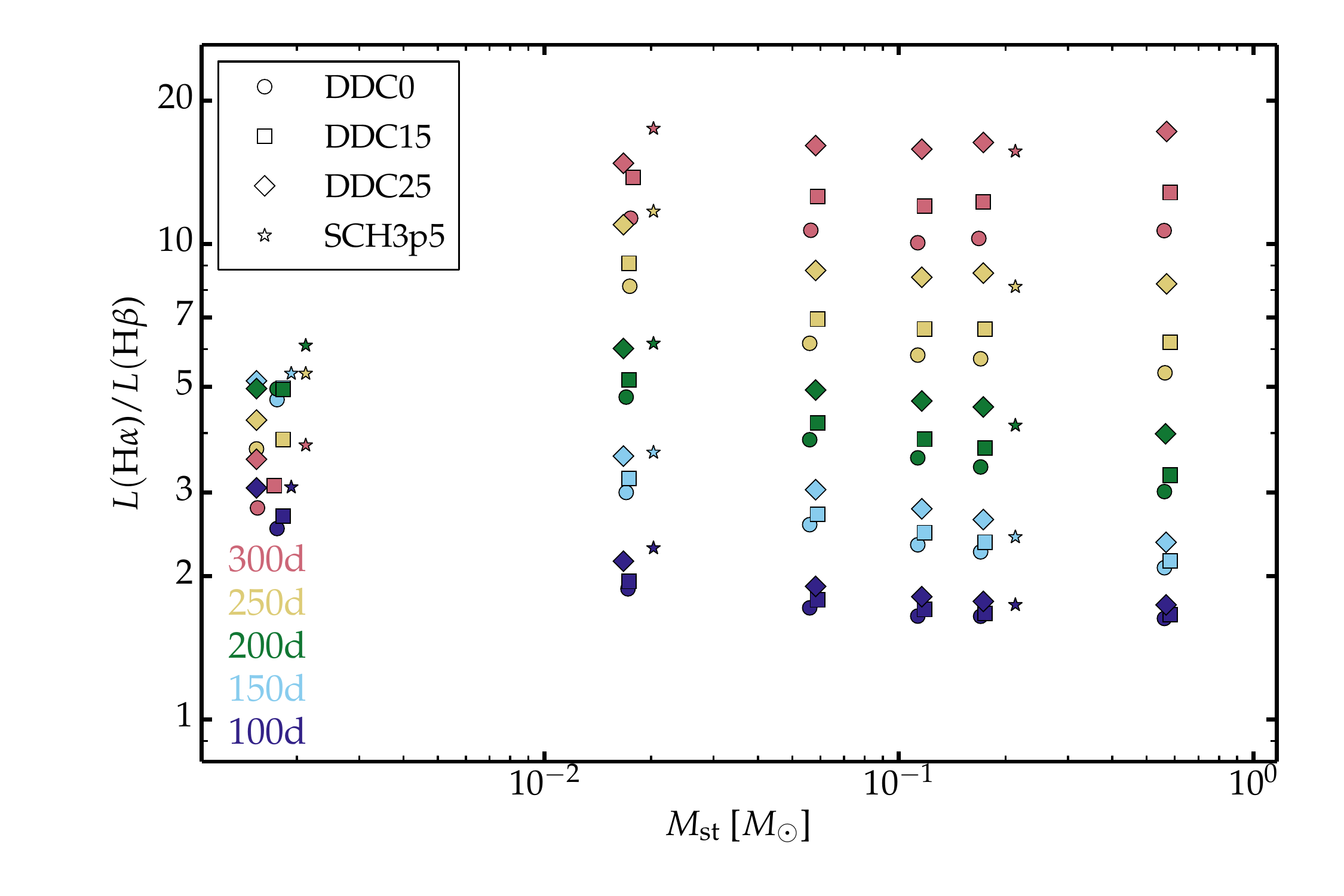}
\vspace{-0.5cm}
\caption{Ratio of the H$\alpha$ to H$\beta$ luminosity for our set of models, adopting only H\one\ lines in the spectrum calculation. Values are scattered, reflecting the influence of intrinsic optical depth effects and the non-thermal process of formation of these lines. This ratio ignores optical depth effects caused by other elements such as iron. Consequently, in the total emergent spectrum (resulting from the contribution of all elements and ions), this ratio is much greater, and potentially infinite.
\label{fig_mh_LHa_div_LHb}
}
\end{figure}

 Figure~\ref{fig_mh_LHa_div_LHb} shows the ratio of H$\alpha$ and H$\beta$ line luminosities computed from the H\one-only spectrum. Apart from cases of very low M$_{\rm st}$, the ratio is not strongly dependent on M$_{\rm st}$ but it varies strongly with time, increasing from about 2 at 100\,d up to $10-20$ at 300\,d. Such a Balmer decrement is much greater than the value of 2.86 for Case B recombination \citep{osterbrock_06} but the conditions here are also very different from those in a photoionized nebula.

A fundamental difference in SNe Ia with stripped material is that the nebula is powered not by ionizing photons from a hot central star but instead from within the ejecta and by radioactive decay. The associated non-thermal effects influence both the excitation and ionization of H and other elements. The process is therefore quite different from photoionization followed by recombination.

The detectability of H$\beta$ in the total emergent spectrum is another issue. As discussed earlier, the blanketing by the metal-rich ejecta affects the Balmer decrement severely in our simulations. Consequently, H$\beta$ is absent or very weak in our simulations of the total emergent spectrum. 

\section{Equivalent width of the H$\alpha$ line}
\label{sect_ew}

Unlike luminosity, the equivalent width (EQW) of an emission line -- or, more
commonly here, an upper limit on its possible strength -- is a quantity that
is directly derivable from an observed SN spectrum. That is, it is independent
of the SN brightness, distance, and extinction, all of which are required to
establish the luminosity of such a line or to place limits on it.  Indeed, for
all but two of the more than $100$ nebular-phase SNe Ia that have so far had
their spectra scrutinized for evidence of stripped material, it is an upper
limit on the EQW of a line that has served as the fundamental, measured
parameter that is derived from the spectrum itself.

Because our models compute the full, nebular-phase SN Ia spectrum in addition
to the expected emergent line luminosities produced by stripped material, they
permit estimates of specific line EQWs to be made.  This can potentially be
used to directly assess -- or to place limits on -- the amount of stripped
material entrained in an SN ejecta without requiring contemporaneous
photometry or estimates of its distance and extinction.

With this in mind, we measured the EQW of the H$\alpha$ line -- hereafter
${\rm EQW (H\alpha)}$ -- in all of our Chandrasekhar-mass model spectra using
a procedure designed to mimic its derivation from an actual SN spectrum.
First, to approximate the ``continuum'' underlying the emission line, points on
the spectrum shortwards and longwards of the H$\alpha$ feature were chosen by
hand and connected with a linear fit.\footnote{This interactive approach was
  chosen over subtracting a model with essentially zero stripped matter (i.e.,
  the model with $M_{\rm st} = 0.000017 M_\odot$) since small amounts of
  ``continuum'' were added by the stripped material  (see Fig.~3) and we
  wished to reproduce, as closely as possible, the procedure applied to an actual
  SN spectrum.  In most cases, the difference was very small.}  The line's
emission profile was then normalized through division by this fitted
continuum. We then calculated ${\rm EQW (H\alpha)}$ by taking the normalized
flux, subtracting one from it, and summing the flux up over the range $6541-6585$\,\AA.  
This range encompasses nearly all of the line emission
(Fig.~\ref{fig_muti_panel_Ha_DDC15}), as it includes $\pm 1000$ \kms\ from the nominal line center.

\begin{figure}[h!]
\includegraphics[width=\hsize]{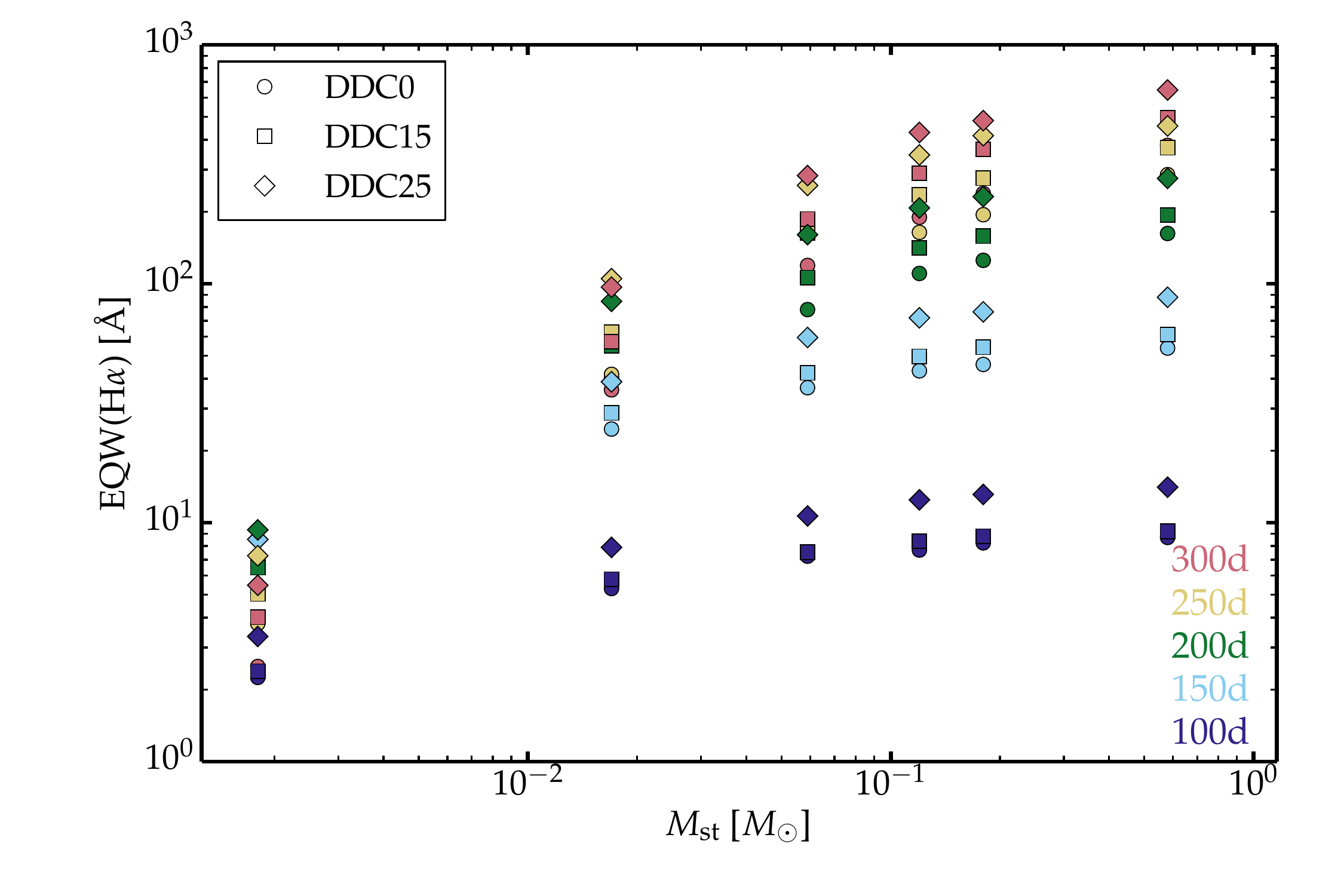}
\vspace{-0.5cm}
\caption{Variation of the H$\alpha$ equivalent width, measured in the
emergent spectrum, as a function of stripped-material mass, $M_{\rm st}$, for
our set of Chandrasekhar-mass delayed detonation models (DDC0, DDC15, and
DDC25).  Color coding defines the SN age.
\label{a16_plot}
}
\end{figure}

We display the results in Fig.~\ref{a16_plot} for all models containing more than $0.001\ M_\odot$; for the two models that contained less than this --  $M_{\rm st} = 1.7 \times 10^{-4}\ M_\odot$ and $M_{\rm st} = 1.7 \times 10^{-5}\ M_\odot$ -- the H$\alpha$ line was very difficult to measure (if apparent at all), and in all measured cases yielded ${\rm EQW (H\alpha)} < 1$\,\AA.  From the figure, three trends are immediately apparent.  First, ${\rm EQW(H\alpha)} $ increases with $M_{\rm st}$ at all epochs and for all models. Second, ${\rm EQW(H\alpha)} $ increases with decreasing \nifs\ mass at all epochs.  This arises primarily because of the reduced light contamination from the SN ejecta for lower \nifs\ mass (this contamination corresponds to the ``pseudo-continuum" by which we normalize the flux when measuring the EQW). Finally, it increases with time for all models with $M_{\rm st} \geq 0.059\ M_\odot$, consistent with our earlier finding that when sufficient stripped material exists it receives an increasing fraction of the decay power relative to the metal-rich ejecta (Sect.~\ref{sect_ddc15}). However, ${\rm EQW (H\alpha)}$ does exhibit more complicated temporal behavior for the two models with lower $M_{\rm st}$. While it generally increases up to day $\sim 200$, it levels off, or even decreases, beyond that time.  As discussed in Sect.~\ref{sect_ha_evol}, this is likely related to the decreasing efficiency of $\gamma$-ray trapping by the stripped material at low $M_{\rm st}$.  We provide analytical fits to the correlations between ${\rm EQW (H\alpha)}$ and $M_{\rm st}$ in Appendix~\ref{appendix_fit_eqw},  along with a suggested prescription for deriving conservative limits on the line's possible strength in observed SN~Ia spectra.

The essential results of this exercise can be summarized succinctly.  First, if a late-time SN~Ia spectrum is obtained with sufficient sensitivity to rule out any H$\alpha$ emission for which ${\rm EQW (H\alpha)}\geq 1$ \AA, a confident limit can be set on the amount of stripped material of $M_{\rm st} < 0.001\ M_\odot$, which is well below the amount expected in the single-degenerate scenario.  Second, balancing the changing effects of relative H$\alpha$ strength, a fading SN and line luminosity, and opacity effects at low $M_{\rm st}$, suggests that the ideal time period to obtain spectra and seek H$\alpha$ emission lies between days $\sim$ 150 and 200.

\section{Conclusions}
\label{sect_conc}

We have presented a grid of radiative transfer calculations for SN Ia ejecta that enclose some stripped material from a non-degenerate companion star. Our set of models covers from faint to luminous SNe Ia, to test the influence of \nifs, as well as \mch\ and sub-\mch\ progenitors to test the influence of the ejecta mass on the results. The ejecta structure is spherically symmetric and parametrized, but uses some constraints from multidimensional hydrodynamics simulations. We extend previous calculations by covering a broader parameter space, by treating the non-LTE radiative-transfer problem in detail. Optical depth effects are treated. The influence of the fast-moving metal-rich ejecta on the radiation emanating from the slower moving stripped material is treated. We cover a range of stripped material masses, from the values of $0.1-0.5$\,\msun\ obtained in multidimensional hydrodynamics simulations down to the low values of about 0.001\,\msun\ that have been inferred from observations.

Our simulations suggest that the emission from the stripped material first appears in the emergent spectrum some time between 50 and 100\,d after explosion for the largest values of $M_{\rm st}$. This delay is wavelength dependent and results from a combination of effects. Metal-line blanketing blocks the radiation from the stripped material in the region below about 5000\,\AA\ as well as in isolated regions (for example over the near infrared Ca\two\ triplet). This prevents the emergence of H$\beta$ photons even at 300\,d (the associated Balmer decrement is thus infinite). In contrast, H$\alpha$ emerges much earlier (for example around 100\,d for $M_{\rm st}\sim$\,0.1\,\msun) because it sits in a spectral region that is relatively free of optically thick metal lines. The same holds for lines like He\one\,1.083\,$\mu$m or O\one\,1.129\,$\mu$m. The second effect is related to the brightness contrast between the emission from the stripped material and that of the overlying metal-rich ejecta. At earlier times, the metal-rich ejecta outshine the stripped material and makes its detection challenging. The problem is less severe for H$\alpha$ because it is intrinsically the strongest optical emission line radiated by the stripped material but also because it is located in a spectral region where the metal-rich ejecta have a moderate brightness. Because of these two effects, H$\alpha$ remains the main observable signature from the stripped material at any ultraviolet, optical, or near-infrared wavelength.

At a given time, the H$\alpha$ luminosity increases with $M_{\rm st}$ because of the greater decay power absorbed by the stripped material (in all our simulations, about 10\% of this power is radiated by H$\alpha$). Being powered by radioactive decay, the H$\alpha$ luminosity generally decreases with time, but this drop is steeper for lower $M_{\rm st}$ because these configurations are less efficient at trapping $\gamma$-rays. Growing $\gamma$-ray escape from the metal-rich ejecta favors the strengthening of H$\alpha$, which becomes the strongest optical line in our simulations at 300\,d for large $M_{\rm st}$. Prior to 200\,d, the H$\alpha$ luminosity can exhibit different behaviors (plateau, rise, or drop), primarily because of optical depth effects. The ${\rm EQW (H\alpha)}$ is found to increase with $M_{\rm st}$, increase with decreasing \nifs\ mass, and generally increase with time except at the lower values of $M_{\rm st}$.

\begin{figure}
\includegraphics[width=\hsize]{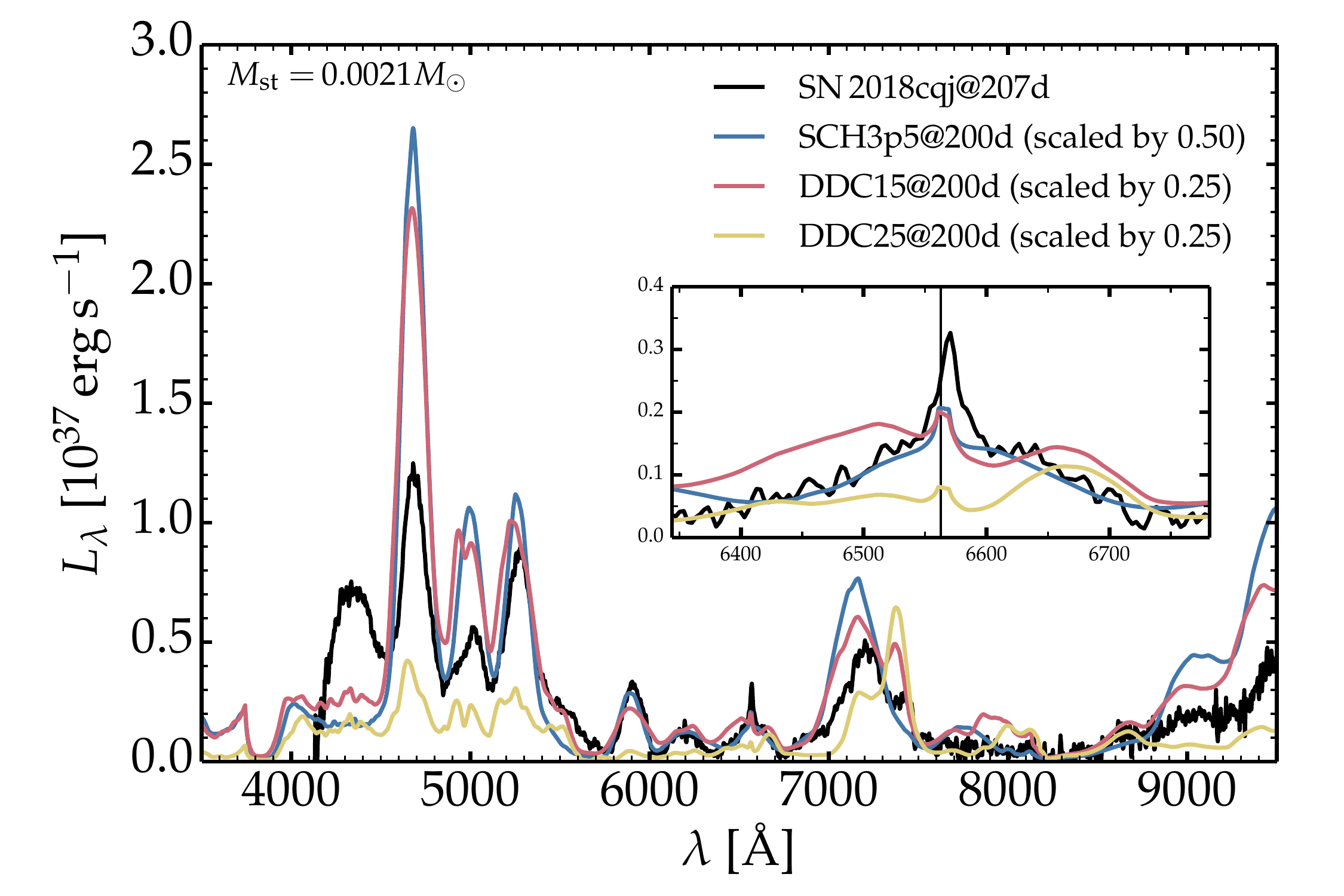}
\vspace{-0.5cm}
\caption{Comparison of the luminosity of model SCH3p5, DDC15, and DDC25 with $M_{\rm st}=0.0021$\,\msun\ at 200\,d to the observations of SN\,2018cqj at 207\,d after explosion. Data as well as SN distance, reddening, and redshift are taken from \citet{prieto_var_ha_19}. The model luminosities are scaled by various amounts (see label).
\label{fig_18cqj}
}
\end{figure}

Optical depth effects significantly impact the radiation from SN Ia ejecta with stripped material. The Balmer lines are intrinsically optically thick even at 300\,d, and emission from the stripped material is also attenuated by the metal-rich ejecta. Optical depth effects caused by the metal-rich ejecta are a fundamental feature of SN Ia with stripped material. These effects are in a large part absent from SN Ia ejecta interacting with CSM since the emission is external to the ejecta rather than deeply embedded within it (optical depth effects influence the receding part of the ejecta but leave intact the front part).

There seems to be a number of ways to distinguish SN Ia ejecta with stripped material and those with CSM interaction. With CSM interaction, H$\alpha$ is observed earlier (it may be observed at any time), H$\beta$ is often detected even at the earliest times (see for example \citealt{hamuy_02ic_03} and \citealt{silverman_ian_13}), the CSM revealed by high-resolution spectroscopy moves slowly (100\,\kms; \citealt{kotak_02ic_04}), and there may be the presence of symmetric electron-scattering wings on H$\alpha$.  With $0.1-0.5$\,\msun\ of stripped material as predicted by hydrodynamical simulations (see for example \citealt{marietta_snia_hyd_00}), Balmer lines cannot be seen much earlier than 100\,d after explosion, H$\alpha$ should strengthen with time relative to the rest of the optical spectrum, H$\beta$ and higher transitions in the series should not be detected during the first year, and H$\alpha$ should be broader but cannot exhibit symmetric electron scattering wings.

Overall, our simulations are in rough agreement with the previous works of \citet{mattila_01el_ha_05} and \citet{botyanszki_ia_neb_sd}. However, our results suggest that optical depth effects produce some important signatures and thus should not be neglected. We also predict weak or absent He\one\ lines for stripped material with a solar composition, with the exception of He\one\,10830\,\AA, in tension with the numerous optical He\one\ lines predicted by \citet{botyanszki_ia_neb_sd}. This may be caused by the different non-thermal and non-LTE treatment, and the neglect of optical depth effects.

The junction between optical and near infrared ranges might also reveal some interesting properties. Since metal-line blanketing is weak in this region, the stripped material can be seen through H\one, He\one, or O\one\ lines. It is not clear whether such lines are predicted in the case of CSM interaction or whether they have ever been observed in SNe Ia with H$\alpha$ detection.

The hydrodynamical models of stripped material from a companion in a SN Ia seem unsuited to explain the observations of SNe Ia with an H$\alpha$ detection (this discrepancy, however, could be resolved by adopting different, and perhaps more physical, initial conditions for the companion, non-degenerate star at the time of explosion; see for example \citealt{justham_snia_11}). Even with the large uncertainties inherent to the radiative transfer calculations, these inferred masses are typically 100 times smaller than predicted by hydrodynamical simulations \citep{kollmeier_18tb_ha_19,prieto_var_ha_19}. Figure~\ref{fig_18cqj} compares one of our calculations with $M_{\rm st}$ of about 0.002\,\msun\ to the observations of SN\,2018cqj at about 200\,d, yielding a good overall match but requiring a very low value for $M_{\rm st}$.

An alternative scenario for the production of H$\alpha$ is interaction with an extended H-rich CSM, so that H$\alpha$ is detected some time after the SN Ia explosion and ceases after a few months once the CSM has been swept up completely by the ejecta. Variations in CSM density structure (or wind mass rate of the progenitor) could yield a wide range of H$\alpha$ luminosities, both in value at a given time and in the evolution of this value until late times. Another possibility, tied to the double-degenerate scenario, is that the inferred 0.001\,\msun\ of H-rich material at low velocity could come from a swept-up giant planet \citep{soker_ia_planets_19}, or from a non-degenerate star in a triple system \citep{thompson_11,kushnir_snia_13,vallely_snian_19}.

The properties of SNe Ia with H$\alpha$ emission reveal an intriguing dichotomy. Events that exhibit strong H$\alpha$ are associated with SNe Ia having a high peak luminosity (or high \nifs\ mass), interacting with a dense H-rich CSM, and are located in relatively young stellar populations. In contrast, the two events that exhibit weak H$\alpha$ emission are associated with SNe Ia with faint peak luminosity (or low \nifs\ mass), have fast-declining light curves, and are located in relatively old stellar populations. There is a dearth of events in between these two extremes. Detectability might play a role here (detecting a weak H$\alpha$ is easier at low \nifs; see Sect.~\ref{sect_ew}), but this is probably not the core reason. A distinct progenitor or explosion scenario may be the cause (see, for example, \citealt{kollmeier_18tb_ha_19}; \citealt{vallely_snian_19}).

A final issue not directly addressed in studies of SNe Ia with H$\alpha$ detection is whether  Chandrasekhar-mass ejecta are suitable to explain the nebular properties of the SN in terms of brightness, color, ionization, line ratios, and other factors. There is indeed growing evidence that the majority of SNe Ia are better explained by the properties of sub-Chandrasekhar-mass ejecta (see, for example, \citealt{kerkwijk_sub_mch_10,kromer_sub_mch_10,kushnir_snia_13,pakmor_snia_13,scalzo_snia_14,blondin_wlr_17,blondin_99by_18,botyanszki_ia_neb_sd,shen_sub_mch_18,flors_neb_snia_19,polin_subch_19,wygoda_snia_t0_19}). In Fig.~\ref{fig_18cqj}, a sub-Chandrasekhar-mass model yields a better match to the observed optical spectrum and brightness than a Chandrasekhar-mass model with a comparable \nifs\ mass. The single-degenerate scenario, intimately tied to the Chandrasekhar-mass for the exploding white dwarf, seems to struggle both in matching the observed H$\alpha$ line strength and the optical radiation from the metal-rich ejecta.

\begin{acknowledgements}

Support for JLP is provided in part by FONDECYT through the grant 1191038 and by the Ministry of Economy, Development, and Tourism's Millennium Science Initiative through grant IC120009, awarded to The Millennium Institute of Astrophysics, MAS. This work was granted access to the HPC resources of  CINES under the allocations  2018 -- A0050410554 and 2019 -- A0070410554 made by GENCI.

\end{acknowledgements}

\appendix

\section{Analytical fits to the correlation between H$\alpha$ luminosity and $M_{\rm st}$}
\label{appendix_fit}

\begin{figure}[h!]
\includegraphics[width=\hsize]{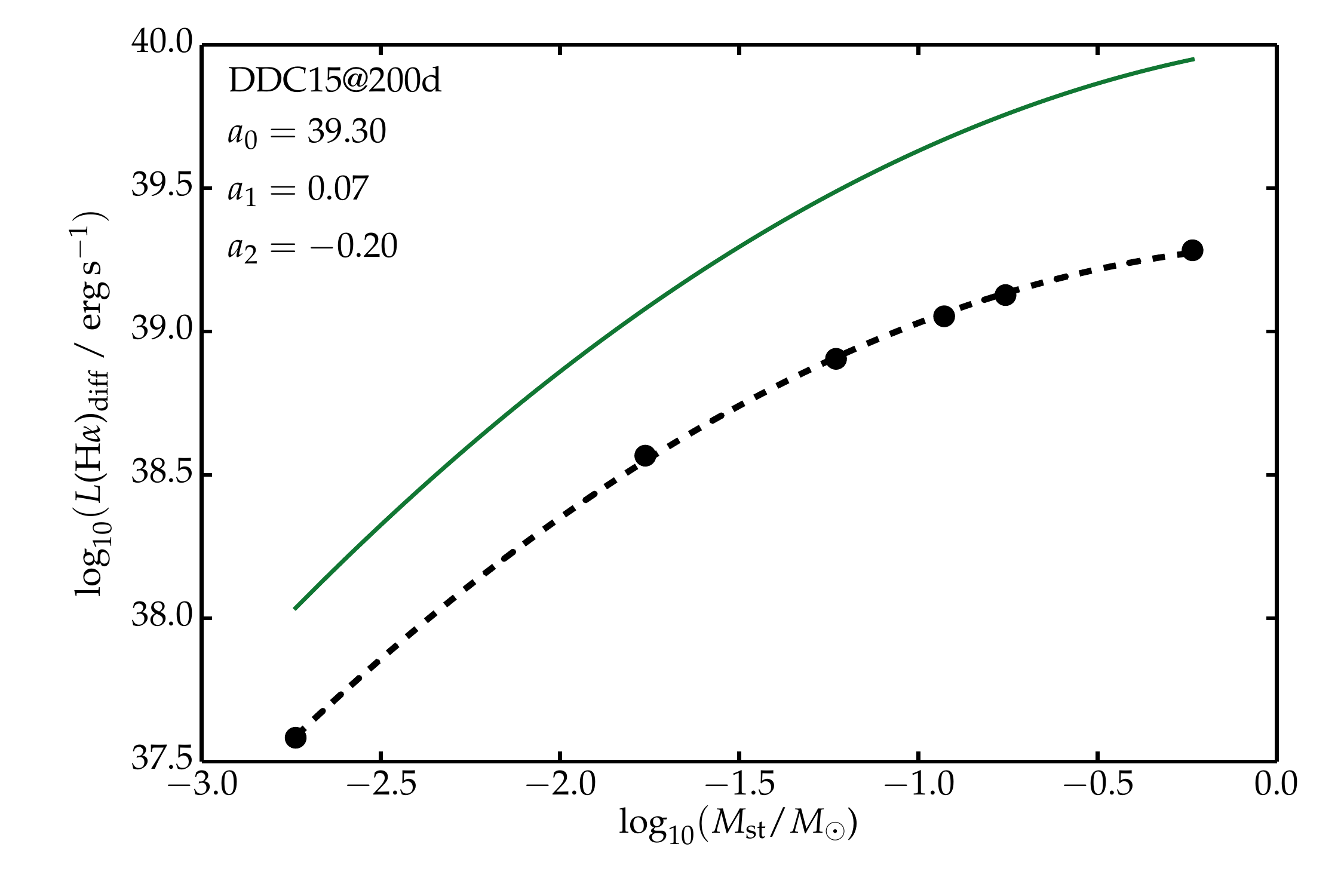}
\caption{Illustration of the polynomial fit (black dashed line) to the distribution of H$\alpha$ luminosity and $M_{\rm st}$ (filled dots) for the model DDC15 at 200\,d, together with the correlation obtained by \citet{botyanszki_ia_neb_sd} for their models at 200\,d. Their Eq.~1 should read  $Y = 40.0 + 0.17 X - 0.2 X^2$, with the nomenclature given in the caption of our Table~\ref{tab_fit} above.
\label{fig_fit}
}
\end{figure}

It is useful to perform polynomial fits to our results for the H$\alpha$ luminosity for the various models, epochs, and stripped material masses. Maintaining the same approach as used in \citet{botyanszki_ia_neb_sd}, we fit a second order polynomial to the distribution of H$\alpha$ luminosity versus $M_{\rm st}$ at each epoch and for each model (which reflects a given \nifs\ mass). The results for the polynomial coefficients are given in Table~\ref{tab_fit} and an illustration of the fit for model DDC15 at 200\,d (which corresponds to the configuration closest to that of \citealt{botyanszki_ia_neb_sd}) is shown in Fig.~\ref{fig_fit}.

\begin{table}[h!]
\caption{Coefficients of the polynomial fits to the correlation between H$\alpha$ luminosity and $M_{\rm st}$ for the models DDC0, DDC15, and DDC25 at 100, 150, 200, 250, and  300\,d. The polynomial has the form $Y = a_0  + a_1 X + a_2 X^2$, where $Y=\log_{10} (L$(H$\alpha$)$_{\rm diff}$ / erg\,s$^{-1})$ and $X=\log_{10} (M_{\rm st} / M_\odot$).
\label{tab_fit}
}
\begin{center}
\begin{tabular}{lcccc}
\hline
Model & Age [d]    & $a_0$  & $a_1$ & $a_2$ \\
\hline
DDC0  &  100  &   39.15  &   -0.01  &   -0.10  \\
DDC0  &  150  &   39.34  &   -0.05  &   -0.17  \\
DDC0  &  200  &   39.35  &    0.04  &   -0.21  \\
DDC0  &  250  &   39.25  &    0.19  &   -0.21  \\
DDC0  &  300  &   39.04  &    0.29  &   -0.21  \\
DDC15  &  100  &   39.11  &   -0.02  &   -0.10  \\
DDC15  &  150  &   39.31  &   -0.02  &   -0.16  \\
DDC15  &  200  &   39.30  &    0.07  &   -0.20  \\
DDC15  &  250  &   39.16  &    0.15  &   -0.24  \\
DDC15  &  300  &   38.97  &    0.30  &   -0.22  \\
DDC25  &  100  &   39.01  &   -0.01  &   -0.09  \\
DDC25  &  150  &   39.22  &    0.06  &   -0.14  \\
DDC25  &  200  &   39.19  &    0.17  &   -0.17  \\
DDC25  &  250  &   39.01  &    0.22  &   -0.21  \\
DDC25  &  300  &   38.79  &    0.37  &   -0.18  \\
\hline
\end{tabular}
\end{center}
\end{table}

\section{${\rm EQW (H\alpha)}$ versus $M_{\rm st}$, and an H$\alpha$ detection-limit prescription}
\label{appendix_fit_eqw}

We have fit polynomials to our results for ${\rm EQW (H\alpha)}$ in a manner
identical to those fitted to its luminosity (see Appendix~A).
Figure~\ref{a16a_plot} provides an illustration of the fit for our standard
explosion model DDC15 at 200 d. The polynomial coefficients for all models are
reported in Table~\ref{tab_fit_eqw}.

\begin{figure}[hb!]
 \includegraphics[width=\hsize]{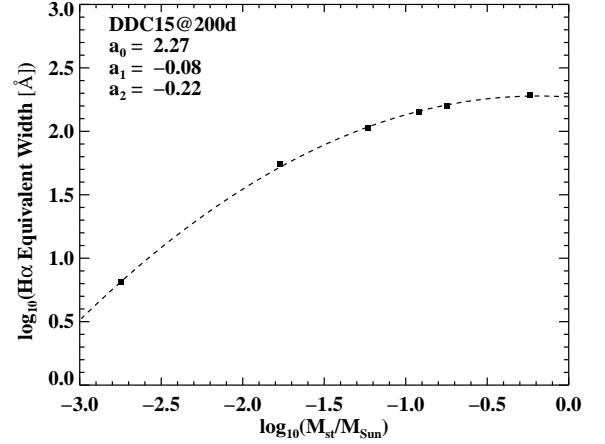}
 \caption{Example of the polynomial fit (black dashed line) to the
 distribution of measured H$\alpha$ equivalent width at simulated values of
 $M_{\rm st}$ for the model DDC15 at 200d.  The caption to
 Table~\ref{tab_fit_eqw} gives the definitions of $a_0$, $a_1$,  $a_2$.
 \label{a16a_plot}
 }
\end{figure}

\begin{table}[h!]
\caption{Coefficients of the polynomial fits to the correlation between
  H$\alpha$ equivalent width and $M_{\rm st}$ for the models DDC0, DDC15, and
  DDC25 at 100, 150, 200, 250, and 300 days. The polynomial has the form $Y =
  a_0 + a_1X + a_2 X^2$, where $Y = \log_{10}({\rm EQW}(H\alpha)/$\AA) and
  $X = \log_{10}(M_{\rm st}/M_\odot)$.  \label{tab_fit_eqw}
}
\begin{center}
\begin{tabular}{lcccc}
\hline
Model & Age [d]     & $a_0$  & $a_1$ & $a_2$ \\
\hline
DDC0  & 100 &  0.93 &  -0.06 &  -0.10\\
DDC0  & 150 &  1.70 &  -0.09 &  -0.16\\
DDC0  & 200 &  2.22 &   0.01 &  -0.20\\
DDC0  & 250 &  2.50 &   0.11 &  -0.21\\
DDC0  & 300 &  2.65 &   0.20 &  -0.23\\
DDC15 & 100 &  0.95 &  -0.07 &  -0.10\\
DDC15 & 150 &  1.76 &  -0.09 &  -0.16\\
DDC15 & 200 &  2.27 &  -0.08 &  -0.22\\
DDC15 & 250 &  2.58 &   0.00 &  -0.25\\
DDC15 & 300 &  2.75 &   0.09 &  -0.25\\
DDC25 & 100 &  1.15 &  -0.01 &  -0.09\\
DDC25 & 150 &  1.92 &  -0.09 &  -0.16\\
DDC25 & 200 &  2.41 &  -0.14 &  -0.24\\
DDC25 & 250 &  2.63 &  -0.20 &  -0.31\\
DDC25 & 300 &  2.81 &  -0.07 &  -0.30\\
\hline
\end{tabular}
\end{center}
\end{table}

Past research tells us that it is nearly always an upper limit, and not a detection,
that is established on the strength of the H$\alpha$ line.  The procedure for
estimating an upper limit on a line's possible strength -- as measured by the
EQW -- has evolved over the years.  The basic methodology was
proposed by \citet{leonard_filippenko_01} based on the seminal work of \citet{hobbs_84}.  The
technique was then implemented specifically for the case of nebular-phase
SNe~Ia by \citet{leonard_07}.  A subsequent empirical investigation by
\citet{sand_snia_ha_18}, in which artificial emission lines were directly injected into
actual SN spectra and then recovered, yielded changes to both the technique and
statistical inferences drawn from it; further refinements were also contributed
by \citet{tucker_snia_ha_19a} based in part on the work of \citet{maguire_snia_ha_16}.  Adopting the accepted, and most conservative, practices from all of the above work
together with the present paper's results yields the following recommended
procedure.

\begin{enumerate}

\item Obtain a high S/N spectrum centered on H$\alpha$ at as
  high a resolution as possible (resolutions of at least $\sim$\,3\,\AA\ are
  desirable to resolve potentially narrow features, since the line width is
  expected to have a viewing-angle dependence) more than 100 days
  post-explosion; epochs between 150 and 200 days are particularly desirable.

\item Remove the redshift and rebin the spectrum to the resolution delivered by
  the spectrograph (as derived, for example, through measurements of widths of
  night-sky lines).

\item Fit a second-order Savitsky-Golay smoothing polynomial \citep{Press92} of
  width $\sim 70$ \AA\ to the spectrum, taking care to exclude the region from
  $6541$ \AA\ --- $6585$ \AA\ to prevent biasing the continuum fit in the event
  H$\alpha$ may be detectable.  Apply a $3\sigma$ clipping to the continuum
  data regions to exclude any artifacts.  If any pixels in the nominal
  H$\alpha$ region are deemed to be untrustworthy (due to, e.g., host-galaxy
  contamination, telluric absorption, or instrumental artifacts), apply the
  pixel-masking technique described by \citet{tucker_snia_ha_19a}.

\item Normalize the spectrum by dividing it by the smoothed continuum, and
  calculate the $1\sigma$ rms fluctuation of the flux around the normalized
  continuum.

\item Difference the smoothed and unsmoothed spectra, and examine the residuals
  for narrow emission near H$\alpha$.  In the event no emission is detected,
  calculate an upper limit on its presence through

\begin{equation}
{\rm EQW (H\alpha)} < 12 \Delta I \sqrt{W_{\rm line} \Delta \lambda}
\label{eqn:1}
,\end{equation}

\noindent where $\Delta I$ is the 1$\sigma$ rms fluctuation of the flux around
the normalized continuum level, $\Delta \lambda$ is the resolution of the
spectrum in \AA\ (which has been set equal to the width of each spectral bin),
$W_{\rm line}$ is the full width at half maximum of the expected feature, typically taken to be $22$
\AA\ ($\sim 1000$ \kms), and ${\rm EQW (H\alpha)}$ is the derived upper bound
on the equivalent width of the undetected H$\alpha$ feature.

\item Use the derived limit on ${\rm EQW (H\alpha)}$ to estimate the maximum
  $M_{\rm st}$ whose effects could remain ``hidden'' in the spectrum.  This may
  be done by using the fits given in Table~\ref{tab_fit_eqw} for the epoch and model
  of choice (i.e., DDC0 for overluminous, DDC15 for normally bright, and DDC25
  for subluminous).  As a simple rule of thumb, if ${\rm EQW (H\alpha)}
  \lesssim 1$ \AA\ at any epoch $> 100$ days, then $M_{\rm st} >
  0.001\ M_{\odot}$ can be confidently ruled out for all models.

\end{enumerate}

\end{document}